%% file: main.tex
\documentclass[11pt]{article}
\usepackage{booktabs}

\usepackage{comment}
\usepackage[utf8]{inputenc}
\usepackage[T1]{fontenc}
\usepackage[a4paper,margin=2.5cm]{geometry}
\usepackage{amsmath,amssymb}
\usepackage{xcolor}
\usepackage[super,numbers,sort&compress]{natbib} 
\usepackage{authblk}
\usepackage{graphicx}
\usepackage[colorlinks=true,linkcolor=blue,citecolor=blue,urlcolor=blue]{hyperref}
\usepackage{placeins} \usepackage{float}

\usepackage{booktabs}
\usepackage{tabularx}
\usepackage{array}

\usepackage{booktabs,array,ragged2e,xltabular,pdflscape,hyperref}
\newcolumntype{L}[1]{>{\RaggedRight\arraybackslash}p{#1}}

\usepackage{pgfplots}
\usepackage{tikz}
\usetikzlibrary{plotmarks}

\pgfplotsset{compat=1.18}

\title{\textbf{Metadevices that compute: mathematical foundations and
experimental implementations for neuromorphic nanophotonics}}

\author[1]{Xian Hadia}
\author[2]{Odysseas Tsilipakos}
\author[2,$\dagger$]{Dimitrios C. Tzarouchis}
\author[1,$\dagger$]{Rapha\"el Pestourie}

\affil[1]{School of Computational Science and Engineering, College of Computing,
Georgia Institute of Technology, Atlanta, GA 30332, USA}
\affil[2]{Theoretical and Physical Chemistry Institute, National Hellenic
Research Foundation (NHRF), 48 Vassileos Constantinou Avenue, 11635 Athens, Greece}

\affil[$\dagger$]{Corresponding authors: \href{mailto:dtzarouchis@eie.gr}{dtzarouchis@eie.gr};
\href{mailto:rpestourie3@gatech.edu}{rpestourie3@gatech.edu}}

\date{Draft of \today}

\begin{document}
\maketitle

\begin{abstract}
\noindent
Wave-based analog computing has re-emerged as a compelling alternative to
digital electronics for specific linear-algebraic and machine-learning
workloads, promising massively parallel, high-bandwidth, and energy-efficient
information processing. Metamaterials, metasurfaces, and integrated
metastructures (collectively, \emph{metadevices}) now perform mathematical
operations, solve integral and differential equations, multiply matrices, and
implement trainable photonic neural networks directly in the wave domain.
This focused review consolidates the mathematical foundations that underpin
such devices, i.e., the mapping between linear operators and wave scattering, the
feedback architectures that realize equation inversion, and the
inverse-design and scientific machine-learning toolchains that make their
synthesis and training tractable, and connects them to the rapidly growing
body of experimental demonstrations, from free-space diffractive processors
to reconfigurable guided-wave metastructures. We conclude with open problems
and opportunities for this emerging neuromorphic nanophotonics ecosystem.
\end{abstract}

\vspace{0.5em}

\section{Introduction}\label{sec:intro}

\begin{figure}[H]
\centering
\includegraphics[width=\textwidth]{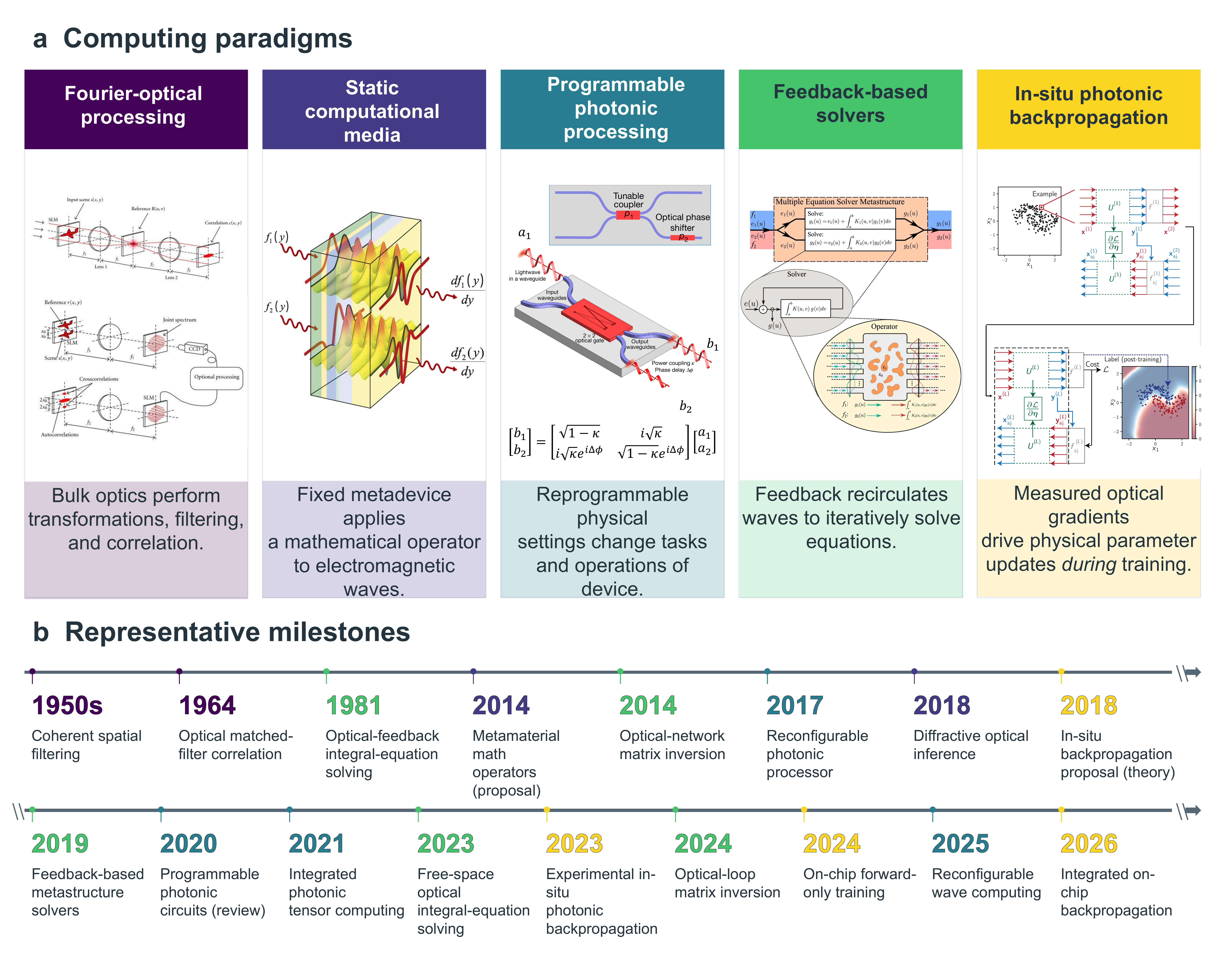}
\caption{
\textbf{From prescribed optical operations to programmable and trainable wave computation.}
\textbf{a}, Representative paradigms: Fourier-optical filtering \cite{ambs2010optical}, fixed computational media \cite{silva2014performing}, programmable photonic circuits \cite{bogaerts2020programmable}, feedback-based equation solving \cite{camacho2021single}, and in-situ backpropagation \cite{pai2023experimentally}. These illustrate complementary capabilities: implementing an operator, changing its parameters, solving through recurrence, and learning from measured gradients.
\textbf{b}, Selected publication milestones, read across the upper row and then the lower row \cite{marechal1953filtre,lugt1964signal,cederquist1981integral,silva2014performing,wu2014computing,perez2017multipurpose,lin2018alloptical,hughes2018training,estakhri2019inverse,bogaerts2020programmable,feldmann2021parallel,pai2023experimentally,cordaro2023solving,chen2024efficient,bandyopadhyay2024single,tzarouchis2025programmable,ashtiani2026integrated}. Dates indicate publication rather than priority.
\emph{Credits:} Fourier-optical schematics reproduced from
ref.~\cite{ambs2010optical} under its Creative Commons
Attribution license; feedback schematic reproduced from
ref.~\cite{camacho2021single} under
\href{https://creativecommons.org/licenses/by/4.0/}{CC BY 4.0}.
Static differentiation and backpropagation images
reproduced/modified with permission from
refs.~\cite{silva2014performing,pai2023experimentally},
AAAS; programmable gate reproduced with minor style changes
from ref.~\cite{bogaerts2020programmable}, with permission
from Springer Nature Customer Service Centre GmbH (SNCSC).
Images were resized and arranged; the timeline is original.
}
\label{fig:timeline}
\end{figure}

The exponential growth of scientific computing, signal processing, and machine-learning workloads is colliding with the energy and scaling limits of digital electronics, thus renewing interest in analog, physics-based computing substrates \cite{mcmahon2023physics,wetzstein2020inference,shastri2021photonics}. Among these, wave-based platforms are particularly promising because superposition, interference, and diffraction naturally enable mathematical operations and signal transformations through passive wave propagation, offering high bandwidth, intrinsic parallelism, and minimal energy expenditure during computation \cite{solli2015analog,zangenehnejad2021analogue}. A seminal development in metamaterial-based mathematical computing was the proposal that suitably engineered metamaterials can perform mathematical operations such as differentiation, integration, and convolution on incident wavefronts \cite{silva2014performing}, followed by demonstrations of inverse-designed metastructures that \textit{solve} integral equations through built-in feedback \cite{estakhri2019inverse}. In parallel, programmable integrated photonics \cite{reck1994experimental,miller2013self,clements2016optimal,shen2017deep} and diffractive optical networks \cite{lin2018alloptical} established that trainable wave systems can implement increasingly complex machine-learning functions. As illustrated in Fig.~\ref{fig:timeline}, these developments trace a broader evolution from fixed optical transformations toward recurrent, reconfigurable, and ultimately trainable physical computing systems. Despite this progress, translating such demonstrations into practical computing platforms remains challenging due to optical loss, analog noise, limited precision, hardware variability, and the overhead associated with signal conversion and control. Additional challenges arise in applications requiring nonlinear operations, dynamic reconfigurability, and trainable architectures. These challenges motivate the co-design of wave-based hardware, computational architectures, and learning algorithms.

\textit{Neuromorphic nanophotonics provides a framework for this co-design by exploiting the physical properties of nanophotonic systems as computational resources.} Rather than treating optical hardware solely as a passive medium for executing prescribed transformations, this approach seeks to integrate wave propagation, interference, feedback, and material responses with learning and adaptive computation \cite{shastri2021photonics,momeni2025training,brunner2025roadmap}. Metamaterials, metasurfaces, and integrated metastructures-collectively referred to here as \textit{metadevices}-offer a particularly versatile platform for engineering these interactions through their geometry, material composition, and electromagnetic coupling. Their computational functionality can be prescribed through inverse design, modified through reconfiguration, or optimized through data-driven training. Although fixed linear operators and equation solvers are not inherently neuromorphic, they provide computational primitives that can be integrated into adaptive, recurrent, and trainable photonic architectures. Their integration with task-dependent learning and appropriate nonlinear responses creates opportunities for more general neuromorphic functionality. These capabilities remain at different stages of experimental maturity, but collectively illustrate a shift from computing with waves toward exploiting their intrinsic dynamics to perform computation directly in the electromagnetic domain.

\begin{figure}[H]
\centering
\includegraphics[width=\textwidth]{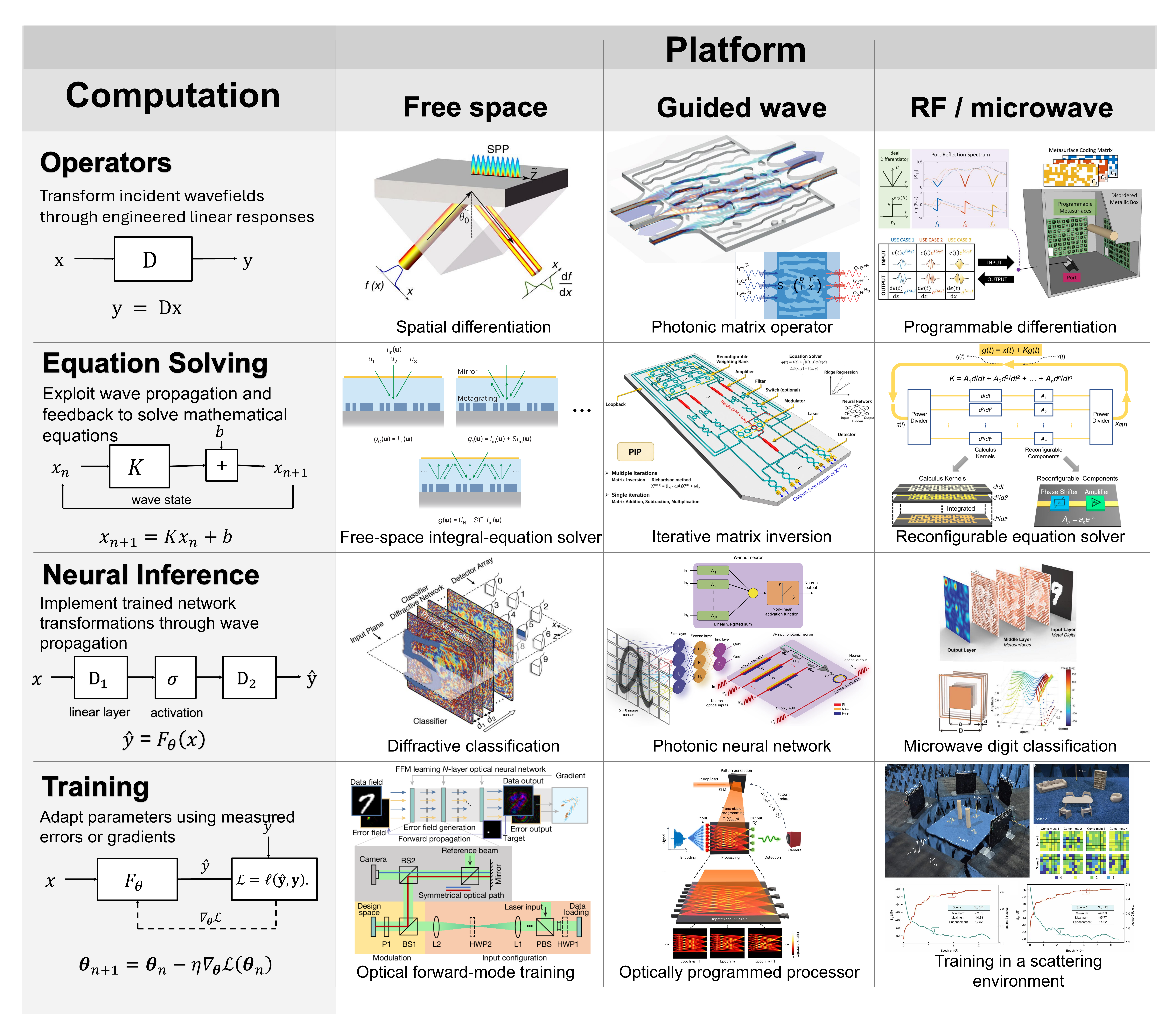}
\caption{
\textbf{Computational functions across wave platforms.}
Rows distinguish computational functions; columns compare free-space, guided-wave, and RF/microwave implementations. Left to right within each row: engineered scattering implements linear operators \cite{zhu2017plasmonic,nikkhah2024inversedesigned,sol2022metaprogrammable}; feedback enables equation solving \cite{cordaro2023solving,chen2024efficient,fu2024reconfigurable}; trained transformations support neural inference \cite{lin2018alloptical,ashtiani2022onchip,gu2024classification}; and learning updates physical parameters using measured responses or gradients \cite{xue2024fully,wu2023lithographyfree,song2026computational}. Examples illustrate functional correspondences across platforms, rather than equivalent architectures or performance.
\emph{Credits:} Images adapted from
refs.~\cite{zhu2017plasmonic,sol2022metaprogrammable,chen2024efficient,fu2024reconfigurable,xue2024fully}
under
\href{https://creativecommons.org/licenses/by/4.0/}{CC BY 4.0};
adapted with permission from
refs.~\cite{nikkhah2024inversedesigned,cordaro2023solving,ashtiani2022onchip,wu2023lithographyfree,song2026computational},
Springer Nature Customer Service Centre GmbH (SNCSC);
and reproduced/modified with permission from
ref.~\cite{lin2018alloptical}, AAAS.
The microwave-classification image is reproduced with
permission from ref.~\cite{gu2024classification},
Copyright 2023 Wiley-VCH GmbH.
Selected panels were cropped, resized, and arranged.
}
\label{fig:taxonomy}
\end{figure}

This evolution motivates a unified mathematical framework connecting electromagnetic wave propagation to computational operations. Beyond mathematical aesthetics, this link formalizes how linear transformations can be physically
realized through engineered scattering processes. Such a framework clarifies which computations arise naturally from wave physics, how these operations can be composed into more sophisticated functions, and where additional mechanisms, including nonlinearity, feedback, and learning, become necessary. Previous reviews have examined the physical limits of optical computing \cite{mcmahon2023physics}, integrated photonic neural networks \cite{shastri2021photonics}, and free-space diffractive computing \cite{hu2024diffractive}. The recent Perspective by Nechipurenko, Gautam, and Shcherbakov examines photonic AI accelerators across cloud and edge platforms, emphasizing scalability, electro-optic interfaces, reconfigurability, and nonlinear responses~\cite{nechipurenko2026photonic}. Our review complements such system-level perspectives by positioning neuromorphic nanophotonics as the co-design of engineered wave operators, recurrent dynamics, and physical learning within computational metadevices. Here, we emphasize the mathematical principles connecting these computational paradigms to their physical realizations. Specifically, we examine how engineered wave scattering implements linear operators, how feedback enables implicit equation solving, and how inverse design and machine learning translate computational objectives into physically realizable electromagnetic structures. We then connect these foundations to experimental metadevices spanning radio-frequency, microwave, and optical regimes, distinguishing fixed mathematical operators from reconfigurable, recurrent, and trainable architectures (Fig.~\ref{fig:taxonomy}).

The review is organized as follows. Sections~\ref{sec:math} and~\ref{sec:sota} establish the mathematical foundations of computing with waves and examine advances in inverse design, physical learning, and computational architectures. Section~\ref{sec:devices} surveys experimental metadevice implementations and relates their demonstrated capabilities to these principles. Finally, Sec.~\ref{sec:outlook} discusses the remaining challenges and opportunities for scalable, programmable, and trainable neuromorphic nanophotonic systems.

\section{Mathematical foundations of computing with waves}\label{sec:math}

\subsection{Wave scattering as a linear operator}

The mathematical foundation of computing with waves begins by treating an electromagnetic field configuration as a single element $\phi \in H$ of a suitable function space. Here, $\phi$ may represent the electric field, the magnetic field, or a state vector containing both. This viewpoint provides a natural language for describing propagation and scattering through operators acting on fields \cite{bossavit1998computational,hiptmair2002finite,reed2012methods}.

Within this framework, let $\theta$ denote the collection of material and geometric design parameters, including, for example, spatially varying permittivity and permeability tensors, conductivity, and dispersive constitutive parameters. At a specified frequency, with fixed boundary conditions and a well-posed linear electromagnetic problem, the structure defines an input-output operator
\begin{equation}
    D_\theta:H_I\longrightarrow H_O,
    \qquad \phi_O=D_\theta\phi_I,
\end{equation}
where $H_I$ and $H_O$ describe admissible input and output field configurations \cite{miller2019waves}. Linearity here is with respect to the incident field at fixed $\theta$; the dependence on the design parameters need not be linear. The effective device or scattering operator is obtained by solving Maxwell's equations for the specified structure and projecting onto the chosen outputs; it is distinct from the differential operator appearing in Maxwell's equations \cite{silva2014performing,miller2019waves}. Free-space propagation, waveguides, lenses, metasurfaces, and complex scattering environments can thus be understood within a common linear-operator framework.

For input and output Hilbert spaces equipped with specified inner products and orthonormal bases $\{\psi_n^{(I)}\}$ and $\{\psi_m^{(O)}\}$, expanding the fields connects the operator description to linear algebra \cite{miller2019waves}. The coefficient vectors $\mathbf a$ and $\mathbf b$ satisfy
\begin{equation}
    \mathbf{b} = \mathbf{D}_\theta \mathbf{a},
    \qquad
    (\mathbf{D}_\theta)_{mn}
    =
    \left\langle
        \psi_m^{(O)},
        D_\theta \psi_n^{(I)}
    \right\rangle_{H_O},
\end{equation}
where each matrix element describes coupling between an input and an output basis function. Retaining finite sets of modes gives a projected, finite-dimensional approximation; exactness requires that the retained output space contain the response to the retained inputs. Electromagnetic modal representations must use the appropriate normalization and duality: radiation spectra may be continuous, and guided or evanescent modes may require power-flux or biorthogonal formulations rather than a simple $L^2$ product \cite{miller2019waves,chew1999waves}. Spatial discretization also permits finite-dimensional input-output maps, but discretizing Maxwell's field equation is a separate step. For example, a discretized linear model $\mathbf A_\theta\mathbf e=\mathbf B\mathbf a$ and output projection $\mathbf b=\mathbf P\mathbf e$ give $\mathbf D_\theta=\mathbf P\mathbf A_\theta^{-1}\mathbf B$ when $\mathbf A_\theta$ is invertible; here $\mathbf B$ injects the excitation and $\mathbf P$ extracts the outputs. Scattering matrices connect port amplitudes, while transfer matrices relate amplitudes on opposite sides of a structure; these representations are related but have different conventions and energy constraints \cite{miller2019waves,chew1999waves,silva2014performing}.

For a finite-dimensional representation, the singular-value decomposition (SVD), $\mathbf D_\theta=\mathbf U\boldsymbol\Sigma\mathbf V^\dagger$, expresses the device map in terms of orthonormal input and output modes coupled through singular-value scaling \cite{miller2000communicating,miller2013self}. This decomposition provides a route to physical implementation: networks of beam splitters and phase shifters can implement arbitrary finite-dimensional unitary transformations, realizing the factors $\mathbf U$ and $\mathbf V^\dagger$, while controlled attenuation implements singular values between zero and one. This construction establishes a direct correspondence between linear algebra and passive wave hardware \cite{reck1994experimental,clements2016optimal}.

Many analog-computing operations admit a natural spatial representation, in which the device operator acts through an integral kernel. For spatially shift-invariant systems, this action reduces to convolution, which becomes multiplication in the spatial-frequency space. Lenses and Fourier optical systems exploit this correspondence to implement filtering, correlation, differentiation, and integration through engineered transfer functions \cite{goodman1969introduction,casasent1976position,lugt1964signal}. More recently, nonlocal metasurfaces have used momentum-dependent scattering responses to realize prescribed spatial-frequency filters in compact nanophotonic hardware, including differentiators and other mathematical operators \cite{kwon2018nonlocal,cordaro2019highindex,zhou2020flat}.

A natural consequence of this operator viewpoint is that wave computation can be interpreted as operator engineering. Once a device is cast as a linear transformation $D_\theta$, designing a photonic structure amounts to choosing physical parameters $\theta$ such that $D_\theta$ approximates the desired operation over specified input and output spaces \cite{silva2014performing,miller2013self}. This perspective provides a common foundation for analog computing with waves, and a conceptual bridge to the equation-solving, inverse-design, and learning frameworks discussed in the following sections.

\subsection{Implicit computation: equation solving with feedback}

Beyond evaluating a prescribed operator, wave computation can solve equations involving that operator. In many applications, the quantity of interest is the solution of a linear system ($\mathbf A\mathbf{x}=\mathbf{b}$), a fixed-point relation ($\mathbf{x}=\mathbf K\mathbf{x}+\mathbf{f}$), or a continuous integral equation. Feedback enables a wave device to address these problems by making the desired solution a steady state of the coupled physical system \cite{estakhri2019inverse,cordaro2023solving,tzarouchis2022mathematical,tzarouchis2025programmable}.

A canonical example is the Fredholm integral equation of the second kind \cite{kress2014linear},
\begin{equation}
    g(u)=f(u)+\int_\Omega k(u,v)g(v)\,dv,
\end{equation}
where $\Omega$ is the integration domain and $(Kg)(u)=\int_\Omega k(u,v)g(v)\,dv$. Assume that $K$ is a bounded operator on the chosen function space. A sufficient condition for the Neumann series to converge in operator norm is $\|K\|<1$; more generally, the condition $\rho(K)<1$ on the spectral radius also ensures convergence~\cite{kress2014linear}. Under this condition,
\begin{equation}
    g=(I-K)^{-1}f=f+Kf+K^2f+\cdots
\end{equation}

A wave system can realize this expansion through a scattering structure implementing $K$ and a feedback path that recirculates the output while adding the input $f$. For a calibrated linear loop with the intended effective operator, successive passes contribute the terms of the Neumann series and the stable steady-state field encodes the solution; physical steady state alone does not establish correctness \cite{estakhri2019inverse,cordaro2023solving}.

This mechanism connects equation-solving metadevices directly to classical iterative numerical methods. The fixed-point update
\begin{equation}
    \mathbf{x}^{(n+1)}=\mathbf K\mathbf{x}^{(n)}+\mathbf{f}
\end{equation}
corresponds to one traversal of the feedback loop. For the finite-dimensional iteration, writing its matrix as $\mathbf K$, the error relative to a fixed point satisfies $\mathbf e^{(n)}=\mathbf K^n\mathbf e^{(0)}$. Thus convergence to a unique fixed point for every initial state and forcing vector is equivalent to $\rho(\mathbf K)<1$. Algebraic invertibility is weaker: $\mathbf I-\mathbf K$ can be invertible even when this iteration diverges. Moreover, nonnormal matrices can transiently amplify errors despite an asymptotic spectral radius below unity. The condition describes an ideal discrete iteration; delays, bandwidth, and additional dynamical elements require a separate stability analysis of the physical loop. Numerical considerations therefore have physical counterparts: loop gain and loss determine the effective iteration operator, conditioning governs sensitivity to errors, and convergence rate determines the required number of recirculations. Preconditioning can improve convergence, while noise and implementation errors limit solution accuracy \cite{saad2003iterative,estakhri2019inverse,tzarouchis2022mathematical}. Early inverse-designed equation-solving metastructures encoded prescribed kernels into passive scattering blocks and used external feedback to realize the corresponding Neumann series. Subsequent implementations extended this approach from guided-wave microwave systems to free-space optical metagratings \cite{estakhri2019inverse,camacho2021single,cordaro2023solving}.

Linear-system solution follows from the same principle; obtaining the full inverse requires solutions for a basis of right-hand sides. Choosing $\mathbf K=\mathbf I-\mathbf A$ and $\mathbf{f}=\mathbf{b}$ gives the steady-state relation $\mathbf A\mathbf{x}=\mathbf{b}$, so that $\mathbf{x}=\mathbf A^{-1}\mathbf{b}$ when $\mathbf A$ is invertible and the feedback iteration converges. Invertibility alone does not ensure $\rho(\mathbf I-\mathbf A)<1$; a scaled or preconditioned iteration may be needed. Reconfigurable metadevice frameworks, including direct-complex-matrix (DCM) architectures, extend this approach to programmable complex-valued kernels and generalized matrix inversions \cite{tzarouchis2022mathematical}. Integrated photonic iterative processors have also demonstrated loop-based matrix inversion and integral- and differential-equation solving with improved input-output efficiency \cite{chen2024efficient}.

Programmable wave-based analog machines further extend this framework to iterations whose operators change during computation, including Newton root finding and constrained inverse-design optimization \cite{tzarouchis2025programmable}. Related photonic iterative approaches address eigenvalue problems through reconfigurable photonic neural-network eigensolvers and shifted inverse iteration \cite{liao2022matrix,klein2024iterative}. Feedback thus extends wave computation from explicit operator evaluation to implicit equation solving, with the solution encoded in a steady state or obtained through a sequence of physical iterations.

\subsection{From design to learning: inverse design and scientific machine learning}

The preceding sections considered how a material distribution $\theta$ determines a device operator $D_\theta$ and how feedback enables equation solving. Inverse design asks the complementary question: how should $\theta$ be chosen to realize a desired operator or computational task? This can be formulated as a constrained optimization problem,
\begin{equation}
    \min_{\theta\in\mathcal C}
    \mathcal L(D_\theta,D_{\mathrm{target}})
    +\lambda R(\theta),
\end{equation}
where $\mathcal L$ measures the discrepancy between realized and desired functionality, $R$ encodes regularization or fabrication objectives, $\lambda\geq0$ weights that term, and $\mathcal C$ specifies admissible designs. The objective function mathematically formalizes the intended engineering functionality, translating physical design goals, such as concentrating optical power at a focal point, realizing a prescribed scattering response, or performing a computational task, into quantities that can be evaluated and optimized \cite{silva2014performing,kwon2018nonlocal,estakhri2019inverse,lin2021endtoend}. A numerical parameterization represents $\theta$ by a finite design vector $\boldsymbol{\theta}=(\theta_1,\ldots,\theta_N)$. By differentiating the objective with respect to these parameters, gradient-based optimization translates prescribed engineering goals into systematic updates of the device geometry or material distribution, thereby automating much of the design process. Maxwell's equations determine the device response, while fabrication, bandwidth, robustness, and material requirements constrain the optimization \cite{jensen2011topology,molesky2018inverse,li2022empowering}. Analog computing devices, equation-solving metastructures, imaging systems, and photonic neural networks therefore share a common design framework.

Adjoint methods make gradient-based inverse design practical even in large design spaces, where an exhaustive exploration is hindered by the curse of dimensionality \cite{bellman2015applied}. For a differentiable scalar objective and a well-posed differentiable forward model, the adjoint method computes gradients with respect to all design variables using at most a single additional electromagnetic solve, rather than one solve per design variable~\cite{lalau2013adjoint}. Multiple frequencies or excitations can increase this count; the cost of each solve and gradient assembly still depends on the discretization and parameterization. Gradients support local optimization but do not guarantee a global optimum or a fixed iteration count. Gradient-based topology optimization has enabled the synthesis of free-form metasurfaces and large-area meta-optics\cite{lalau2013adjoint,jensen2011topology,pestourie2018inverse,molesky2018inverse,pestourie2020assume}. As the devices grow, simulation and optimization costs remain a bottleneck~\cite{pestourie2025fast}, which motivates reduced-order models, hierarchical parameterizations, and efficient optimization strategies \cite{christiansen2021inverse,su2020nanophotonic,li2022empowering,marzban2026inverse,huang2026neural}.

\begin{figure}[htbp]
    \centering
    \includegraphics[width=16cm]{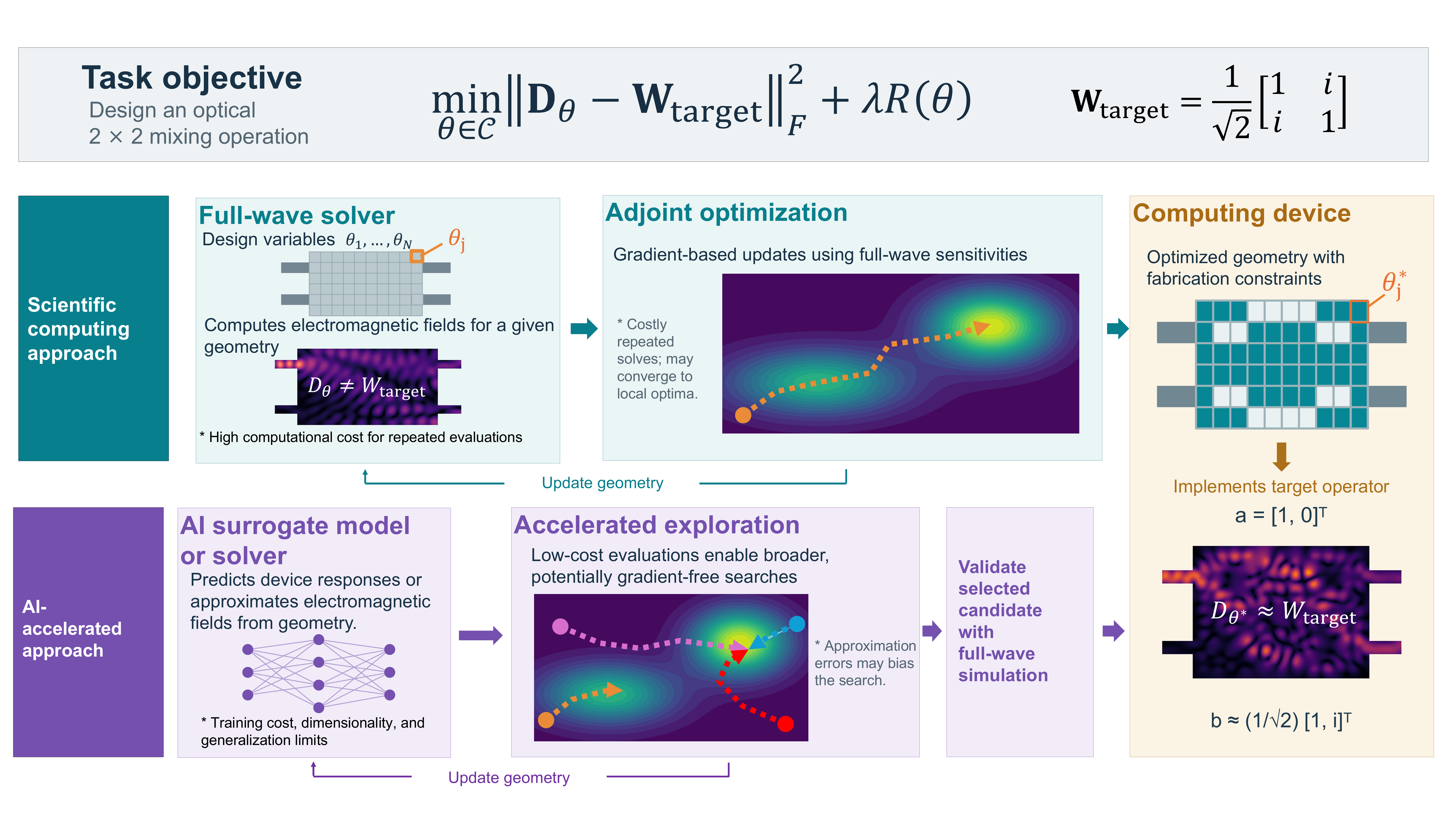}
\caption{
\textbf{Inverse design translates a target operator into a physical device.}
Design parameters $\boldsymbol{\theta}$ are optimized to approximate a target $2\times2$ optical operator by minimizing $\|\mathbf D(\boldsymbol{\theta})-\mathbf W_{\mathrm{target}}\|_F^2+\lambda R(\boldsymbol{\theta})$ over the fabrication-constrained set $\mathcal C$, where $R$ is a design regularizer and $\lambda$ its weight. Full-wave simulation supplies electromagnetic responses and adjoint gradients \cite{lalau2013adjoint}; an AI surrogate model or accelerated solver reduces evaluation cost during exploration, with selected designs validated by full-wave simulation \cite{pestourie2020active,pestourie2023physics,lu2022multifidelity}. Here, $\theta_i$ denotes an individual design parameter and stars denote optimized values. The diagram is original; field maps and optimization landscapes are illustrative.
}
    \label{fig:inverse_design}
\end{figure}

Scientific machine learning complements these methods by reducing the cost of repeated response evaluations. Surrogate models, active learning, and multifidelity approaches approximate mappings between design parameters and device responses while selectively using high-fidelity simulations \cite{pestourie2020active,pestourie2023physics,lu2022multifidelity}. For a neural-network surrogate $S_{\boldsymbol{\eta}}(\boldsymbol{\theta})$, $\boldsymbol{\eta}$ denotes the trainable network weights and biases\footnote{as a reminder, $\boldsymbol{\theta}$ represents the physical design parameters supplied as inputs}. Training adjusts $\boldsymbol{\eta}$ to approximate simulated or measured responses; subsequent surrogate-assisted inverse design holds $\boldsymbol{\eta}$ fixed and differentiates the predicted objective with respect to $\boldsymbol{\theta}$ to guide design updates. This enables rapid exploration of the design space without requiring a full-wave simulation at every optimization iteration. Neural operators learn mappings between function spaces, including parameter-dependent PDE solution operators, and provide a complementary class of surrogate models \cite{lu2021learning,li2020fourier,kovachki2023neural}. Surrogate accuracy and gradients must be validated against the governing model, particularly outside the training distribution; physics-informed losses alone do not guarantee that Maxwell's equations or conservation laws are satisfied exactly~\cite{lu2021physics}. Combined with differentiable optimization, these tools support task-level optical and computational co-design \cite{sitzmann2018endtoend,lin2021endtoend}. Complementary work develops differentiable device parameterizations \cite{colburn2021inverse}, metasurface inverse-design methods \cite{li2022inverse}, and extended-depth-of-focus meta-optics for computational imaging \cite{bayati2022inverse}. Such objectives can include downstream inference accuracy and privacy, allowing the optical encoding itself to be optimized for the information required by the task \cite{kang2026vlmaware,tasneem2026privacy}. However, the computational cost of generating training data must be amortized through repeated surrogate evaluations for this approach to provide a net computational advantage~\cite{pestourie2025fast}. Figure~\ref{fig:inverse_design} illustrates a tutorial-like workflow combining electromagnetic simulation, surrogate learning, and fabrication-aware optimization.

\subsection{Computing-with-waves as neural networks}

Neural networks extend the operator viewpoint from implementing individual
transformations to composing trainable linear transformations with nonlinear activations,
or learning the dynamics and readout of recurrent architectures. A
feed-forward wave neural network may be written schematically as
\begin{equation}
    \mathbf{x}_{\ell+1}
    =
    \sigma_\ell\left(\mathbf D_{\theta_\ell}\mathbf{x}_\ell+\mathbf c_\ell\right),
\end{equation}
where $\mathbf{x}_\ell$ denotes the state at layer $\ell$,
$\theta_\ell$ collects the physical design or control parameters of layer $\ell$ (for example, geometry, refractive indices, or phase-shifter settings), $\mathbf D_{\theta_\ell}$ is the resulting scattering matrix whose entries are the effective linear weights in the chosen modal basis, $\theta_\ell$ is distinct from the surrogate neural-network weights and biases $\boldsymbol{\eta}$ defined above,
$\mathbf c_\ell$ is an optional bias supplied by an auxiliary coherent input or electronic stage,
and $\sigma_\ell$ denotes a nonlinear transformation between
successive layers, which may be realized through material nonlinearities,
optoelectronic conversion, or measurement-and-feedback processes
\cite{shen2017deep,lin2018alloptical,wetzstein2020inference,
shastri2021photonics}. This schematic describes architectures with interlayer activations. Passive diffractive networks without such elements remain linear in field amplitude, even when their final intensity readout is nonlinear \cite{lin2018alloptical}.

Wave systems naturally support high-dimensional linear mixing and, in resonant or feedback-based architectures, dynamical state evolution. These capabilities support feed-forward linear processing and recurrent physical neural networks, respectively; nonlinear transformations must be supplied where required by the computational architecture. In feed-forward architectures, propagation through diffractive layers, interferometer meshes, or metastructures implements matrix-vector multiplication and modal transformations. In resonant, delayed, or feedback-based systems, the internal field itself serves as a dynamical hidden state, allowing the physical evolution to be interpreted as a recurrent neural network, a reservoir computer, or a continuous-time dynamical system. Computation therefore emerges directly from the native wave dynamics rather than being imposed on top of them \cite{hughes2019wave,marcucci2020theory,wright2022deep}.

A fundamental limitation arises, however, because linear wave propagation remains linear regardless of the complexity of the underlying structure. Composing fixed linear scattering operators without intervening nonlinear transformations produces another linear operator, meaning that depth alone does not generate the expressive hierarchy associated with modern neural networks. Nonlinear transformations, including nonlinear state dependence or nonlinear measurement and feedback, must therefore be introduced to increase representational power. Existing approaches include demonstrated optoelectronic activations \cite{ashtiani2022onchip}, and material nonlinearities or nonlinear detection-and-re-injection schemes surveyed in \cite{shastri2021photonics}. Phase-change media can instead provide stored linear weights, as in the photonic tensor core of \cite{feldmann2021parallel}; weight storage alone is not an activation. Effective nonlinear maps can also arise through the encoding of computational inputs. Wanjura and Marquardt encode inputs in system parameters: although the probe-to-output map is linear at fixed parameters, the scattering matrix depends nonlinearly on the encoded input \cite{wanjura2024fully}. Yildirim and colleagues repeatedly encode data in modulation layers, making the output nonlinear in the encoded data while propagation remains linear in the optical field \cite{yildirim2024nonlinear}. Such schemes differ from nonlinear hidden-state dynamics: nonlinear encoding or intensity readout can supply a nonlinear input-output map while the internal field evolution remains linear. Linear interference or stable linear recurrence alone does not violate superposition.

Unlike digital neural networks, wave neural networks use physical fields and device responses and therefore inherit the constraints of the underlying hardware. Noise, loss, finite detector dynamic range, fabrication disorder, calibration drift, quantized tuning, and input-output conversion overhead all limit achievable precision and scalability. Consequently, wave neural networks should be viewed not as exact floating-point implementations of neural computation but as analog physical approximators whose performance depends on hardware-aware training, robustness, and task tolerance to error. Relevant metrics, therefore, extend beyond model accuracy to include latency, throughput, energy efficiency, bandwidth, footprint, and the cost of interfacing between digital and wave domains \cite{mcmahon2023physics,momeni2025training,abouhamdan2025programmable,brunner2025roadmap}.

When referring to neuromorphic wave computing, we point at architectures in which physical wave transformations or dynamics implement learned neural computations. Fixed operators, programmable equation solvers, and reconfigurable processors are enabling components rather than automatically neural systems. Diffractive networks, integrated photonic neural networks, reservoir computers, and trainable metadevices can nevertheless be analyzed within a common framework that separates linear field processing, input encoding, nonlinear activation or readout, and task-dependent learning \cite{hughes2019wave,wright2022deep,momeni2025training}. We summarize these correspondences in Table~\ref{tab:dictionary}.

\begin{table}[htpb]
    \centering
    \small
    \caption{\textbf{Correspondences between computational concepts and wave physics.}
    Fields encode computational states, propagation and scattering implement
    linear transformations, feedback enables implicit equation solving, and
    optimization adjusts physical parameters to realize a target computation.
    The final column identifies representative formulations and implementations
    of each correspondence; these are illustrative rather than claims of priority.}
    \label{tab:dictionary}
    \renewcommand{\arraystretch}{1.35}
    \setlength{\tabcolsep}{5pt}

    \begin{tabular}{
        >{\raggedright\arraybackslash}p{0.24\linewidth}
        >{\raggedright\arraybackslash}p{0.27\linewidth}
        >{\raggedright\arraybackslash}p{0.42\linewidth}
    }
        \toprule
        \textbf{Computation} &
        \textbf{Wave physics} &
        \textbf{Correspondence and representative works} \\
        \midrule

        Vector / computational state
        \newline
        ($\mathbf a$, $\mathbf b$, $\mathbf x$)
        &
        Field configuration
        \newline
        ($\phi\in H$) and its basis coefficients
        &
        Modal expansion maps fields to coefficient vectors \cite{miller2019waves}. \\ \specialrule{0.2pt}{5pt}{5pt}

        Linear operator / matrix-vector product
        \newline
        ($\mathbf b=\mathbf D_\theta\mathbf a$)
        &
        Propagation or scattering
        \newline
        ($\phi_O=D_\theta\phi_I$)
        &
        Engineered scattering implements mathematical operators and vector-matrix products \cite{silva2014performing,nikkhah2024inversedesigned}. \\ \specialrule{0.2pt}{5pt}{5pt}

        Singular-value decomposition
        \newline
        ($\mathbf D=\mathbf U\boldsymbol{\Sigma}\mathbf V^\dagger$)
        &
        Orthogonal input-output channels and mode conversion
        &
        Orthogonal channels define coupling strengths; interferometric meshes realize unitary factors \cite{miller2000communicating,reck1994experimental,
        clements2016optimal}. \\ \specialrule{0.2pt}{5pt}{5pt}

        Spatial filtering
        \newline
        ($\widehat g(k)=T(k)\widehat f(k)$)
        &
        Spatial-frequency-dependent transmission or scattering
        &
        Spatial-frequency filters implement correlation and nonlocal optical processing \cite{lugt1964signal,kwon2018nonlocal}. \\ \specialrule{0.2pt}{5pt}{5pt}

        Linear solve / fixed point
        \newline
        ($g=f+Kg$)
        \newline
        ($g=(I-K)^{-1}f$)
        &
        Steady state of a device coupled to feedback
        &
        Scattering encodes the kernel; feedback yields an integral-equation solution \cite{estakhri2019inverse,cordaro2023solving}. \\ \specialrule{0.2pt}{5pt}{5pt}

        Iterative update
        \newline
        ($\mathbf x^{(n+1)}
        =\mathbf K\mathbf x^{(n)}+\mathbf f$)
        &
        Repeated scattering and input addition
        &
        Wave feedback implements stationary and non-stationary iterations. For the linear fixed-point update, convergence from arbitrary
        initial states requires $\rho(\mathbf K)<1$
        \cite{tzarouchis2022mathematical,
        tzarouchis2025programmable}. \\ \specialrule{0.2pt}{5pt}{5pt}

        Inverse design / parameter training
        \newline
        ($\theta^*=\operatorname*{arg\,min}_{\theta\in\mathcal C}
        \mathcal J(\theta)$)
        &
        Optimization of geometry, materials, or control settings
        &
        Adjoint simulations or in-situ optical measurements provide design gradients \cite{lalau2013adjoint,hughes2018training,
        pai2023experimentally}. \\ \specialrule{0.2pt}{5pt}{5pt}

        Feed-forward neural layer
        \newline
        ($\mathbf x_{\ell+1}
        =\sigma_\ell(\mathbf D_{\theta_\ell}\mathbf x_\ell)$)
        &
        Trainable optical mixing with an activation or readout
        &
        Interferometric or diffractive mixing forms neural layers; nonlinear activations or intensity readout complete the mapping \cite{shen2017deep,lin2018alloptical,
        ashtiani2022onchip}. \\ \specialrule{0.2pt}{5pt}{5pt}

        Recurrent state update
        \newline
        ($\mathbf h_{n+1}
        =F_\theta(\mathbf h_n,\mathbf u_n)$)
        &
        Temporal wave evolution with internal memory
        &
        Field evolution implements recurrent states; material parameters set trainable couplings \cite{hughes2019wave}. \\

        \bottomrule
    \end{tabular}
\end{table} 

\section{State of the art: algorithms, training, and benchmarks}\label{sec:sota}

\subsection{From fixed optical transformations to programmable wave computation}

Section~\ref{sec:math} established a common framework for wave systems as operators, feedback solvers, and inverse-designed computational devices. Current developments build on these foundations through programmable optical transformations, equation-solving architectures \cite{estakhri2019inverse,cordaro2023solving}, and photonic implementations of machine-learning models. Depending on the task, geometry, material properties, optical control settings, feedback paths, or readout rules are designed or adjusted to achieve the desired computational function \cite{wetzstein2020inference,shastri2021photonics,mcmahon2023physics,brunner2025roadmap}. Early photonic neural-network demonstrations established that optical propagation can naturally implement the linear layers of neural networks. Coherent integrated photonic circuits use interferometric meshes to realize programmable matrix-vector multiplication, while diffractive free-space systems implement learned transformations through cascaded phase or amplitude masks \cite{shen2017deep,lin2018alloptical}. These demonstrations proved photonics to be a very compelling platform for high-throughput linear algebra since weighted superposition and modal mixing are natural consequences of wave propagation. Simultaneously, the central difficulty within wave computing is also exposed: useful neural computation requires training, \textit{nonlinear} activation, calibration, robustness, memory, and efficient input-output interfacing. A physical neural network implements its computational mapping through physical dynamics; this does not require training to occur on the device. Its parameters may be optimized in silico using a model of the physical system, adjusted in situ using hardware measurements, or trained through a hybrid procedure \cite{wright2022deep,momeni2025training}. These developments broaden the role of photonics from implementing individual optical transformations to realizing programmable computational systems whose physical responses, control schemes, and readout are designed together for a specified task.

\subsection{Training physical and photonic neural networks}
Implementing a digital operator in its physical analogue introduces
imperfections, drift, noise, and finite precision. Training may take place entirely in silico, but the optimized model must represent the physical device sufficiently accurately for its parameters to transfer to hardware. The
learning procedure must account for how optical weights are represented, how
nonlinearities are implemented, how gradients are obtained, and how model
parameters are transferred to or updated on a physical platform
\cite{momeni2023backpropagation,momeni2025training}. A common approach
is digital-twin training, in which a differentiable model of the optical
system is optimized in software and the resulting parameters are fabricated
or programmed into the device
\cite{lin2018alloptical,zhou2021large,mcmahon2023physics,hammami2026expressivity}. This strategy leverages mature machine-learning tools and works well when the optical model is accurate, but its performance can degrade when the realized hardware deviates from the simulation. In-situ training addresses this model-mismatch problem by measuring gradients or perturbation responses directly on the physical platform. In coherent photonic circuits, reciprocity and interference can be used to obtain gradient information from forward and adjoint optical fields, allowing the hardware that performs inference to also participate in backpropagation \cite{hughes2018training,pai2023experimentally,ashtiani2026integrated}. A complementary direction is forward-only or backpropagation-free learning, where update directions are inferred from perturbations, local measurements, equilibrium shifts, or feedback signals during ordinary device operation \cite{momeni2023backpropagation,xue2024fully,bandyopadhyay2024single,momeni2025training}. These methods may be less sample-efficient than idealized digital backpropagation, but they are often more compatible with analog systems that lack exact adjoint access or stable component models. Together, these paradigms suggest that fabrication disorder, analog noise, and drift should not only be corrected after deployment, but incorporated into the learned physical model.

\subsection{Large-scale photonic inference engines and chiplets}
The field is also moving from proof-of-principle classifiers toward large-scale processors with many optical channels, multiplexed wavelengths, programmable weights, and system-level integration. Integrated photonic tensor cores, on-chip photonic neural networks, all-analog photoelectronic computing chips, and photonic chiplets have demonstrated increasingly complex inference workloads by exploiting wavelength, spatial, temporal, and modal parallelism \cite{feldmann2021parallel,ashtiani2022onchip,chen2023allanalog,xu2024taichi}. These systems demonstrate the central architectural opportunity of photonic computing: a large number of multiply-accumulate-like operations can be mapped onto passive propagation, interference, diffraction, or electro-optic modulation rather than sequential digital arithmetic.

Different platforms scale in different ways. Integrated photonic processors provide stronger compatibility with electronic control, packaging, and reconfigurable weights, but face challenges associated with insertion loss, tuning power, footprint, and calibration. Free-space and diffractive systems scale through aperture area, spatial pixels, and meta-atom count, making them attractive for image processing and optical front-end computation where the data already arrive as spatially distributed fields \cite{lin2018alloptical,zhou2021large,hu2024diffractive}. Hybrid chiplet architectures such as Taichi combine diffractive propagation and integrated interference to distribute computation across multiple optical modules, illustrating a broader shift from single-device demonstrations toward system-level photonic processors \cite{xu2024taichi}. Across these examples, the key lesson is that photonic scaling will likely not come from directly copying digital neural-network architectures into optics, but from co-designing optical physics, electronic control, and digital post-processing so each subsystem performs the computation for which it is physically best suited.

\subsection{Programmability as a key enabler}

Programmability is the key distinction between a compelling optical demonstration and a useful computing platform. Fixed diffractive masks, static inverse-designed structures, and passive metasurfaces can perform impressive task-specific transformations, but adaptive AI hardware requires reconfigurable weights, nonvolatile memory, calibration strategies, and hardware-aware training \cite{bogaerts2020programmable,abouhamdan2025programmable,brunner2025roadmap}. This is especially important for neuromorphic nanophotonics, where the target is not merely to fabricate one optical function, but to create physical systems that can be trained, updated, corrected, and reused.

Programmability can be introduced at several levels. Interferometric meshes tune optical phases and amplitudes through thermo-optic, electro-optic, carrier-based, or other integrated phase-shifting mechanisms, making them a natural platform for field-programmable photonic circuits \cite{bogaerts2020programmable}. Phase-change materials provide nonvolatile, multilevel optical weights for photonic memory and in-memory computing, enabling programmed matrix-vector operations without continuous static tuning power \cite{feldmann2021parallel}. In free-space systems, spatial light modulators and digital micromirror devices enable large-scale pixel-wise programmability, albeit often with slower refresh rates, larger optical footprints, and optoelectronic control overhead \cite{zhou2021large,pierangeli2019large,hu2024diffractive}. Programmable metasurfaces and reconfigurable scattering media provide another route, replacing networks of discrete components with compact media whose scattering responses are electronically, optically, or materially controlled \cite{abouhamdan2025programmable,tzarouchis2025programmable}. The common challenge is that programmability introduces overhead: tuning energy, reconfiguration and control latency, calibration time, drift, crosstalk, limited precision, and control circuitry can dominate the performance of the optical core. As a result, the most promising architectures are not simply those with the highest raw optical throughput, but those whose programmability, stability, and system integration allow the optical computation to remain useful after fabrication.

\subsection{Benchmarks, limits, and metrics}
The central benchmarking question for wave-based processors has shifted from ``can light compute'' to ``under what conditions does wave hardware provide a system-level advantage over digital computing?'' While photonic accelerators offer compelling advantages for high-bandwidth linear algebra, analog signal processing, optimization, and neural inference, digital hardware continues to dominate in programmability, precision, memory integration, and general-purpose computing \cite{mcmahon2023physics,brunner2025roadmap}.

\begin{figure}[htbp]
    \centering
    \includegraphics[width=\textwidth]{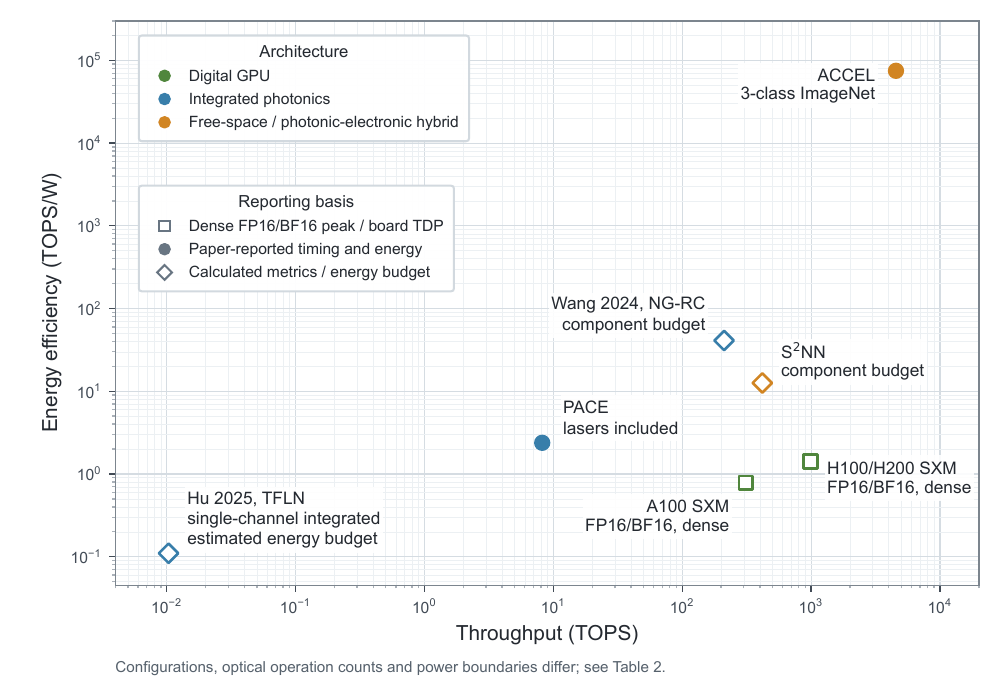}
    \caption{\textbf{Performance landscape with explicit reporting bases.}
    Energy efficiency versus throughput on shared logarithmic axes; the companion Table~\ref{tab:fig4_benchmarks} gives the values and source-specific assumptions. Green open squares show dense FP16/BF16 Tensor Core peak throughput divided by rated board power for individual A100, H100, and H200 SXM GPUs\cite{nvidiaA100specs2026,nvidiaH100specs2026,nvidia2024h200}. Dense peaks are half the advertised sparse peaks; one multiplication and one addition count as two floating-point operations. H100 and H200 overlap exactly and share a label. Filled circles show paper-reported timing and energy for PACE, including lasers\cite{hua2025integrated}, and ACCEL, using the Methods values for three-class ImageNet\cite{chen2023allanalog}. Open diamonds identify calculated energy/component budgets: S$^2$NN's published equivalent-operation count and 33.2~W budget\cite{xu2026sparse}, Hu et al.'s demonstrated single-channel laser-integrated TFLN circuit (10.33~GOPS), with an estimated 9.015~pJ/operation energy budget corresponding to approximately 0.111~TOPS/W (plotted at the source-rounded 0.11~TOPS/W)\cite{hu2025integrated}, and Wang et al.'s silicon next-generation reservoir engine\cite{wang2024ultrafast}. Blue denotes integrated photonics; orange denotes free-space diffractive or photonic-electronic hybrid hardware. Digital floating-point operations and source-defined optical equivalent operations are displayed in TOPS, but optical arithmetic precision, workloads, operation counts, and peripheral boundaries differ. Channel counts and configurations are retained as documented in each source; Hu et al.'s separate 31-channel configuration is not used. Calculated component budgets are distinguished from paper-reported timing and energy and do not establish measured full-apparatus efficiency. The plot is a qualified literature landscape, not a normalized performance ranking.}
    \label{fig:performance_landscape}
\end{figure}

Meaningful comparisons, however, require careful attention to benchmarking methodology. Reported throughput, latency, and energy efficiency often correspond to different computational primitives, precision formats, and measurement boundaries, with some works reporting chip-only performance while others include optical sources, electronic control, or complete system overheads \cite{mcmahon2023physics,hua2025integrated,abouhamdan2025programmable}. Figure~\ref{fig:performance_landscape} retains a single set of axes but distinguishes reporting bases with marker shape; Table~\ref{tab:fig4_benchmarks} records the numerical values and assumptions. The digital references share a common convention: dense FP16/BF16 Tensor Core peak throughput for one SXM GPU, divided by its rated board power. These ratios are not measured application efficiencies and exclude host and facility overhead. The optical points retain source-supported coordinate pairs with their reported illumination terms, but use each paper's own operation counts, workloads, and power accounting. Paper-reported timing and energy are distinguished from a component-power-budget calculation. Common axes therefore do not establish common arithmetic precision, equivalent workloads, or identical peripheral boundaries, and no inter-paper speedup or energy-efficiency ratio is inferred. A fully normalized comparison would require the same workload and accuracy target, a consistent operation-count convention, and measured energy over the same system boundary.

This distinction is particularly important for computational metasurfaces and equation-solving metadevices. Many demonstrations validate mathematical operators, integral-equation solvers, and programmable matrix inversion without reporting throughput and energy efficiency on a common system-level basis\cite{silva2014performing,estakhri2019inverse,cordaro2023solving,tzarouchis2022mathematical,tzarouchis2025programmable}. They are therefore excluded from Fig.~\ref{fig:performance_landscape}. Propagation delays or energy estimates inferred from dimensions and operating frequencies are useful physical proxies, but do not substitute for measured end-to-end performance including illumination, detection, conversion, and control.

Overall, the benchmark landscape reflects differences in reporting methodologies, hardware performance, and technological maturity. Vendor peak specifications provide a reproducible digital reference, while optical demonstrations establish performance for particular tasks and accounting conventions. Up-to-date, many metadevice architectures remain proof-of-concept demonstrations focused primarily on validating new computational mechanisms. We argue that whether future programmable metadevices can ultimately achieve an advantage on a matched system-level benchmark remains an important open research question. For the moment, the lack of a unified benchmarking approach for such proof-of-concept devices makes it hard to create a comparative scatter plot.

\newpage

\input{figures/fig4_benchmarks.tex}

\FloatBarrier

\section{Metadevices and experimental implementations}\label{sec:devices}

The decade since the proposal of computational metamaterials\cite{silva2014performing} has produced a remarkably diverse experimental record: computing structures have been demonstrated in free space and on chip; at optical, terahertz, microwave, and even acoustic frequencies; in passive and field-programmable forms; and at scales ranging from single-operator metasurfaces to processors executing full neural-network workloads\cite{zangenehnejad2021analogue,wetzstein2020inference,mcmahon2023physics}. This section surveys these implementations through the lens of the \emph{metadevice}-an engineered, subwavelength-structured medium whose scattering response \emph{is} the computation-organized by architecture (Secs.~\ref{sec:freespace}-\ref{sec:implicit}), by physical platform (Sec.~\ref{sec:rf}), and by enabling technology (Sec.~\ref{sec:materials}). We then assess what the experiments collectively demonstrate (Sec.~\ref{sec:assessment}) and close with the opportunity space that, in our view, metamaterials, metasurfaces, and metadevices uniquely open (Sec.~\ref{sec:beyond}).

\subsection{Computing metasurfaces: free-space analog operators}\label{sec:freespace}
The most direct route from metamaterial to mathematics is the spatial filter: because free-space propagation diagonalizes linear space-invariant operators in the transverse wavevector, a surface whose angular response $H(k_x,k_y)$ is engineered to match a target transfer function applies that operator to every point of an incident wavefront simultaneously. The original metamaterial-computing concept proposed two routes: implementing an operator through its Green's function in a thin slab, and synthesizing its transfer function in the Fourier plane of a lens system\cite{silva2014performing}. Related experimental spatial differentiators are discussed below. Plasmonic interference at a single interface provided an experimental plasmonic spatial differentiator, performing edge detection directly on optical images\cite{zhu2017plasmonic}; engineered nonlocality, i.e., a deliberately tailored angular dispersion near guided-mode resonances, was then identified as a general design principle for metasurface-based optical signal processing\cite{kwon2018nonlocal} and used to design first- and second-order dielectric differentiators, with second-order image differentiation demonstrated experimentally\cite{cordaro2019highindex}. Isotropic two-dimensional Laplacian operators followed, with a photonic-crystal-slab differentiator placed directly in an imaging path to deliver all-optical edge extraction and phase visualization\cite{zhou2020flat}.

These devices crystallize the defining strengths of the explicit, single-pass metadevice: the computation is performed passively, in parallel across millions of spatial pixels, without active power consumption in the passive operator itself and with an optical latency set by propagation; illumination, detection, and readout still require energy. Their equally instructive limitations-one fixed kernel per fabricated structure, a finite numerical aperture and bandwidth over which the synthesized $H(k_x,k_y)$ remains faithful, and polarization sensitivity-delineate precisely the challenges (programmability, cascadability, fidelity) that the architectures of Secs.~\ref{sec:diffractive}-\ref{sec:implicit} address. In application terms, computing metasurfaces are already compelling as passive optical front ends that compress or pre-process scenes (edge extraction, phase contrast, feature projection) before a single photon is detected, reducing the downstream digital load of machine-vision and microscopy pipelines\cite{zhou2020flat,zangenehnejad2021analogue}.

For instance, silicon metasurfaces have experimentally demonstrated second-order spatial differentiation at $\lambda=726$~nm with numerical aperture up to $0.35$\cite{cordaro2019highindex}. Similarly, photonic-crystal-slab differentiators have demonstrated two-dimensional image differentiation with numerical aperture up to $0.32$\cite{zhou2020flat}. Transfer-function transmission at selected angles should be distinguished from the total optical throughput of an image-processing system.

\subsection{Diffractive processors: from passive classifiers to programmable optoelectronic engines}\label{sec:diffractive}

Where computing metasurfaces implement a single prescribed operator, cascaded diffractive layers implement \emph{learned}, task-specific transformations. Diffractive deep neural networks (DNNs) established the paradigm: multilayer phase masks, trained in silico and 3D-printed for terahertz operation, classified handwritten digits all-optically at the speed of light, with no active power consumed by the passive layers themselves, excluding illumination, detection, and readout\cite{lin2018alloptical}. The family has since expanded to universal linear transformations, quantitative phase imaging, polarization- and wavelength-multiplexed processing, and task-specific optical front ends, together with quantitative design rules for accuracy, diffraction efficiency, and robustness, as surveyed in our recent account of free-space diffractive computing\cite{hu2024diffractive}.

DNNs are essentially layered feed-forward networks that consist of a series of structured surfaces through which photons diffract. They have emerged as promising all-optical neuromorphic processors \cite{Li2024}, since they allow for a much higher density of neurons and easier connectivity (synapses) in space, compared to their integrated counterparts. Moreover, DNNs are naturally suited for tasks such as all-optical image analysis, feature detection, and object classification, without the need for any pre-processing of the input. The advantages of DNNs are especially highlighted in compact, metasurface-based implementations. The trainable parameters are the complex-valued transmission coefficients of the unit cells. However, in most cases thus far metasurface-based DNNs are non-reconfigurable and can implement a single inference task \cite{Wu2019}, or two when parallelism in terms of the two orthogonal linear polarization channels is exploited \cite{luo2022metasurface}. The learning process is implemented offline on a computer: the transmission coefficients of the unit cells are derived and the resulting surfaces are fabricated and assembled. In contrast, incorporating a tuning mechanism in the unit cell can allow for on-site programmability as well as the capability of reconfiguring the processor to perform various inference tasks.

Two complementary directions have lifted the paradigm from fixed classifiers to genuine processors. The first is \emph{programmability}. Replacing printed layers with spatial light modulators yielded the reconfigurable diffractive processing unit, an optoelectronic architecture supporting millions of neurons and various network topologies, with adaptive in-system training that mitigates accumulated experimental errors, as demonstrated for high-speed image and video recognition \cite{zhou2021large}. The same step was taken at microwave frequencies with a multilayer array of 1-bit digital-coding transmissive metasurfaces: a programmable diffractive deep neural network that performs inference directly on microwave wavefronts and can be re-trained in situ for new tasks\cite{liu2022programmable} using field-programmable gate arrays. 

The second direction is \emph{hybridization and scale}. The all-analog chip ACCEL\footnote{reporting a computing speed of 4.6 Peta-OPS with over 99\% performed optically, a systemic energy efficiency of 74.8 Peta-OPS/W, and a frame processing latency of 72~ns\cite{chen2023allanalog}} couples a diffractive optical front end to analog electronic computing, eliminating analog-to-digital conversion between layers and reporting orders-of-magnitude energy-efficiency and speed gains over the compared GPUs for the reported vision tasks, accuracy levels, equivalent-operation counts, and power-accounting boundaries\cite{chen2023allanalog}, while the photonic chiplet Taichi distributes diffractive and interferometric computing across modules, reaching million-parameter network capacity, thousand-category classification, and an energy efficiency of 160~TOPS/W\cite{xu2024taichi}. At the opposite extreme of the energy axis, free-space optical neural networks have operated in the quantum-noise-limited regime using fewer than one photon per scalar multiplication, demonstrating task-specific inference at low detected photon budgets per multiplication, rather than a universal energy floor for complete computing systems\cite{wang2022optical}. Diffractive hardware also computes beyond feed-forward inference: spatial photonic Ising machines encode thousands of spins on a modulator and anneal combinatorial-optimization problems through measurement-feedback dynamics\cite{pierangeli2019large}.

Taken together, free-space diffractive systems are today the scaling champions of wave computing-parallelism grows with aperture area rather than with on-chip component count-at the price of bulkier optics, limited (or optoelectronically mediated) nonlinearity, and an analog precision that must be actively managed by error-aware training\cite{hu2024diffractive,zhou2021large}.
Efficient nonlinear activation is key to introducing ``depth'' in the transformation solution space and enhance the convergence speed and inference ability \cite{Ryou2021}. A promising candidate for nonlinear activation is saturable absorption, since even ultrathin 2D materials with fast response (sub-ps carrier lifetime) and low saturation intensity (below 10~MW/cm$^2$) are available \cite{Christopoulos2025}. These material metrics motivate an activation element but do not by themselves demonstrate neural-network activation or establish system-level switching energy, insertion loss, or cascadability. Another important challenge in multilayer deep free-space neural processors is to reduce the overall thickness. This is mainly determined by the interlayer distance, which is dictated by diffraction of light scattered by the meta-atoms in free space. A disruptive approach to significantly decrease this distance can be based on the concept of ``squeezed space'' \cite{Guo2020,Chen2021}, where a much thinner dielectric structure is made to emulate the diffraction properties of free space.

\subsection{Integrated photonic processors: meshes, metastructures, and on-chip networks}\label{sec:integrated}
On chip, the canonical programmable linear processor is the Mach--Zehnder interferometer mesh, rooted in constructive decompositions of unitary operators \cite{reck1994experimental,clements2016optimal} and subsequently developed into programmable integrated processors for optical linear algebra and neural-network inference \cite{miller2013self,shen2017deep}; the broader technology of field-programmable photonic circuits is reviewed by Bogaerts \textit{et al.}\cite{bogaerts2020programmable}.
Reconfigurable photonic circuits have also been developed for partial-differential-equation solving by Ye and colleagues, including Juejun Hu \cite{ye2024reconfigurable}. Meshes carried photonic neural inference into the laboratory-a 56-interferometer processor performing vowel recognition\cite{shen2017deep}-and have been proposed for solving differential equations through suitably closed recursions\cite{tzarouchis2020solving}. More recently they have closed the loop on \emph{learning} itself: in situ backpropagation, with gradients measured interferometrically on the very hardware that computes the forward pass\cite{pai2023experimentally}; fully forward-mode training that dispenses with reciprocal back-injection\cite{xue2024fully}; a single chip co-integrating coherent matrix algebra, optoelectronic nonlinearities, and forward-only on-chip training\cite{bandyopadhyay2024single}; and, most recently, complete on-chip backpropagation training of an integrated photonic neural network\cite{ashtiani2026integrated}. End-to-end on-chip inference with optoelectronic activation has reached sub-nanosecond ($\sim$0.57~ns) classification latency\cite{ashtiani2022onchip}, and in-memory photonic tensor cores combining phase-change weights with soliton-comb wavelength multiplexing execute parallel convolutions at rates of trillions of multiply-accumulate operations per second\cite{feldmann2021parallel}.

The metadevice perspective enters where component-by-component architectures meet their footprint and loss limits: inverse-designed, free-form metastructures collapse entire operator blocks into wavelength-scale regions. On silicon photonics, low-index-contrast inverse-designed structures now perform vector-matrix multiplication directly within the propagation of guided waves, trading the $O(N^2)$ tiling of discrete interferometers for a single computational medium compatible with foundry processing\cite{nikkhah2024inversedesigned}; related metastructures compute prescribed forward-scattering operators, perform several operations in parallel within one structure, and can be paired with reconfigurable couplers for post-fabrication trimming\cite{camacho2021single,nikkhah2022inverse}. Industrial-scale integration has meanwhile demonstrated that photonic linear algebra survives contact with real workloads: a packaged accelerator with more than 16{,}000 photonic components on a single chip delivers ultralow-latency computing that outpaces a commercial GPU on latency-critical tasks\cite{hua2025integrated}, and 128$\times$128 photonic tensor cores have executed ResNet, BERT, and deep-reinforcement-learning models at near-electronic precision\cite{ahmed2025universal}. These milestones, achieved largely with conventional component libraries, define the performance envelope that metastructure-based computing cores must now enter-with the promise of doing so at a fraction of the area and insertion loss.

\subsection{Implicit computing: equation-solving metastructures and programmable metadevice machines}\label{sec:implicit}
The feed-forward operator blocks discussed above compute \emph{explicitly}: one pass, one operator. Closing a feedback loop around a metadevice turns it into an \emph{implicit} computer whose steady state solves the intended equation when the calibrated linear loop implements the prescribed operator and is stable-physically realizing the stationary iterations formalized in Sec.~\ref{sec:math}. Inverse-designed metastructures embedded in microwave waveguide networks provided the founding demonstration, solving Fredholm integral equations of the second kind with the kernel encoded in a dielectric scattering block and the recursion implemented by an external feedback path\cite{estakhri2019inverse}; single structures performing several such operations in parallel followed\cite{camacho2021single}. The concept has since left the waveguide: ultrathin inverse-designed metagratings with built-in feedback solve integral equations in free space at optical wavelengths, merging the equation-solving paradigm with flat optics\cite{cordaro2023solving}.

Reconfigurability-arguably the central experimental challenge for equation-solving hardware-motivated the direct-complex-matrix (DCM) architecture, in which an $N\times N$ complex-valued kernel is mapped one-to-one onto tunable elements, avoiding a priori matrix decompositions and their associated calibration chains, and enabling general mathematical operations, matrix inversion, and root finding on the same hardware\cite{tzarouchis2022mathematical}. Its experimental realization at radio frequencies demonstrates a programmable radio-frequency direct-complex-matrix computing \emph{machine}: arbitrary complex kernels are written element-by-element, linear systems are solved in the physics of the structure, and-closing a conceptual loop-the machine has been operated inside an optimization loop to inverse design other metastructures, ``a metastructure that designs metastructures''\cite{tzarouchis2025programmable}. Operating at a carrier frequency of $45$~MHz, the realized $5 \times 5$ complex-matrix hardware incorporates 25 active, microcontroller-driven multiplier channels offering $>360^\circ$ phase control and a $\sim 47$~dB dynamic range ($-30$~dB to $+17$~dB)\cite{tzarouchis2025programmable}. In open-loop operation, the device executes matrix-vector multiplications within an optical/wave response time of $\sim 110$~ns ($5$ wave periods), while in closed-loop feedback mode it achieves real-time stationary linear-system solution ($A^{-1}\mathbf{b}$) and non-stationary optimization, such as root finding via Newton's method and inverse design via Lagrangian multipliers\cite{tzarouchis2025programmable}. 

Programmable scattering environments offer a complementary, decomposition-free route to the same end: a chaotic cavity dressed with a tunable metasurface has been shaped into an over-the-air analog differentiator that processes several signals on distinct carriers in parallel and is reprogrammable in both operator order and carrier\cite{sol2022metaprogrammable}, after the same class of platform had been used to perform reconfigurable matrix-vector products-including discrete Fourier transforms-directly on indoor wireless signals\cite{delhougne2018leveraging}.

\begin{figure*}[t]
\centering
\includegraphics[width=\textwidth]{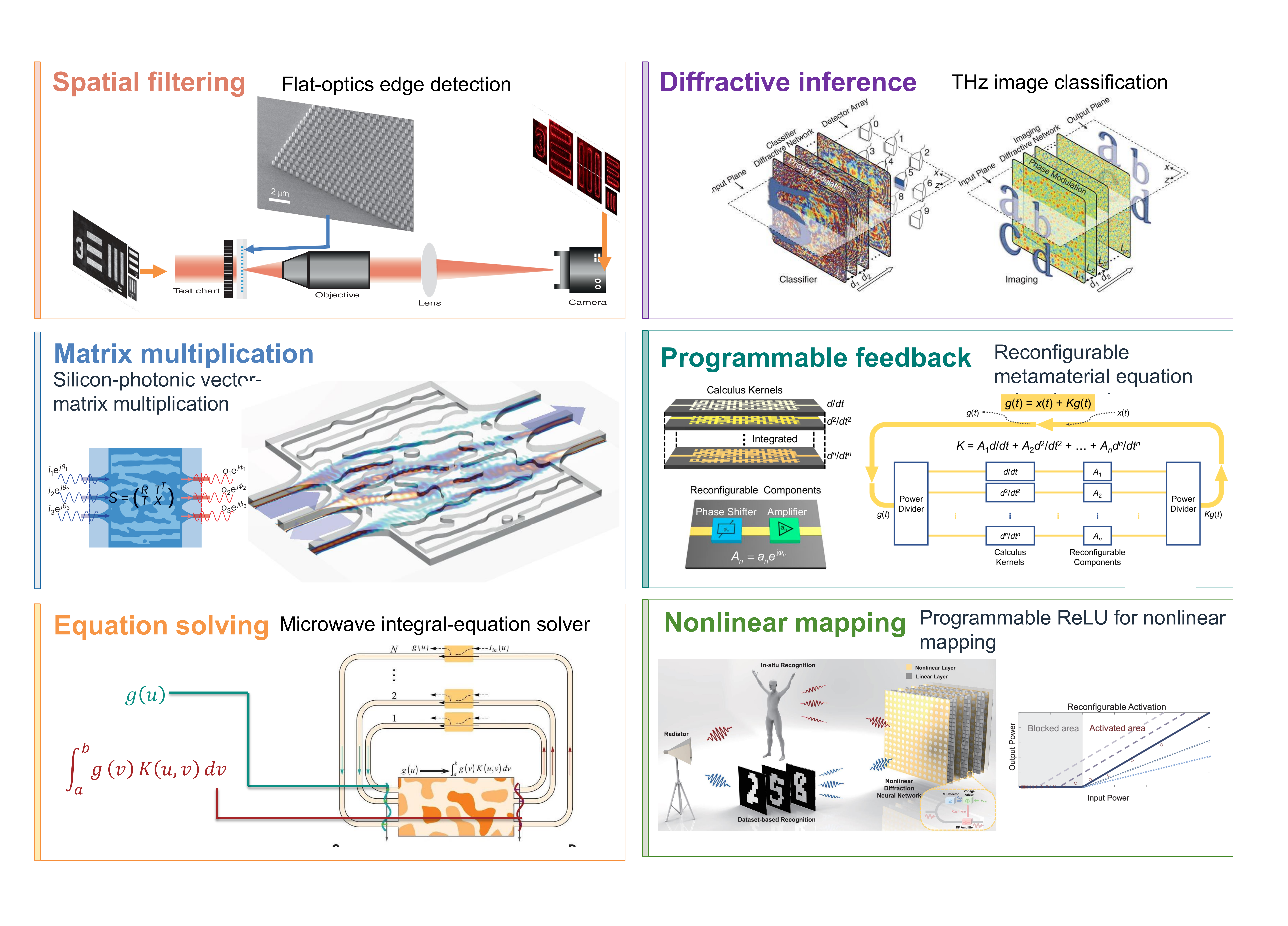}
\caption{
\textbf{Physical mechanisms underlying experimental wave computation.}
Left column, top to bottom: spatial filtering extracts image edges \cite{zhou2020flat}, guided-wave scattering implements vector-matrix multiplication \cite{nikkhah2024inversedesigned}, and microwave feedback solves an integral equation \cite{estakhri2019inverse}. Right column, top to bottom: trained diffractive layers perform classification and imaging \cite{lin2018alloptical}, reconfigurable feedback changes the equation implemented by the device \cite{fu2024reconfigurable}, and programmable nonlinear activation extends diffractive processing beyond linear field transformations \cite{ning2025multilayer}.
\emph{Credits:} Top left and middle left adapted,
and bottom right reproduced, with permission from
refs.~\cite{zhou2020flat} (top left), \cite{nikkhah2024inversedesigned} (middle left), and \cite{ning2025multilayer} (bottom right),
respectively, Springer Nature Customer Service Centre
GmbH (SNCSC).
Top right and bottom left reproduced/modified with permission
from refs.~\cite{lin2018alloptical} (top right) and \cite{estakhri2019inverse} (bottom left),
respectively, AAAS.
Middle right adapted from ref.~\cite{fu2024reconfigurable}
under
\href{https://creativecommons.org/licenses/by/4.0/}{CC BY 4.0};
selected panels were cropped and arranged.
}
\label{fig:computing_metadevices}
\end{figure*}

Implicit metadevices remain small ($N$ of order units to tens) but are qualitatively distinctive: they compute $A^{-1}\mathbf b$ for a specified right-hand side in a field relaxation; obtaining the full inverse requires independent right-hand sides. Convergence time depends on round-trip delay, loop spectrum, conditioning, stability, and required accuracy, and they map naturally onto the fixed-point and deep-equilibrium layers of modern implicit machine learning, a bridge we develop in Sec.~\ref{sec:beyond}. Representative experimentally realized devices spanning these computing
functions are shown in Fig.~\ref{fig:computing_metadevices}.

\subsection{Beyond optics: microwave information metasurfaces, plasmonic networks, and acoustic-mechanical computing}\label{sec:rf}
Because the underlying mathematics is wave-agnostic, every metadevice concept has a counterpart across the spectrum-and the microwave domain, with its mature tunable components and direct line to wireless systems, has often led on programmability. Digital-coding and programmable metasurfaces, which discretize the local reflection or transmission response into electronically switchable states\cite{cui2014coding}, supplied the substrate for the programmable diffractive networks of Sec.~\ref{sec:diffractive}\cite{liu2022programmable} and for software-defined apertures with embedded networks of controller chips, conceived as autonomously communicating metasurface fabrics\cite{tsilipakos2020toward}. The same hardware merges sensing with computing: intelligent metasurface systems can jointly optimize programmable microwave measurements and downstream inference, treating the reconfigurable physical layer itself as part of the learned sensing pipeline \cite{delhougne2020learned}. Related microwave imager-recognizer architectures combine learned illumination patterns with neural post-processing to image scenes and recognize human gestures in situ at microwave frequencies \cite{li2019intelligent}-an early embodiment of computing-integrated sensing. Guided microwave platforms have followed: spoof-surface-plasmon networks with varactor-weighted couplings and detector-amplifier activation cells realize fully programmable neural networks, including 4$\times$4 matrix-multiplying supercells and a 16-input partially connected classifier, operating directly on microwave signals\cite{gao2023programmable}.

Acoustics and mechanics complete the picture and stress-test the concepts under strong disorder and intrinsic nonlinearity. Topological acoustic metamaterials perform spatial differentiation with built-in robustness to geometric imperfections\cite{zangenehnejad2019topological}; trained acoustic and other physical systems were among the first deep \emph{physical} neural networks, learned either through digital twins and physics-aware backpropagation\cite{hughes2019wave,wright2022deep} or entirely without backpropagation through physics-local learning rules\cite{momeni2023backpropagation}; and architected mechanical lattices of tunable-stiffness beams learn target behaviors directly in their structure, pointing toward materials that are themselves trainable\cite{lee2022mechanical}. For the telecommunications-minded reader, the message of this subsection is pointed: the component base of reconfigurable intelligent surfaces and smart radio environments\cite{direnzo2020smart} is the same component base as that of programmable computing metasurfaces-an overlap whose consequences we take up in Sec.~\ref{sec:beyond}.

\subsection{Materials and technologies for reconfigurability}\label{sec:materials}
\begin{figure*}[t]
\centering
\includegraphics[width=0.95\textwidth]{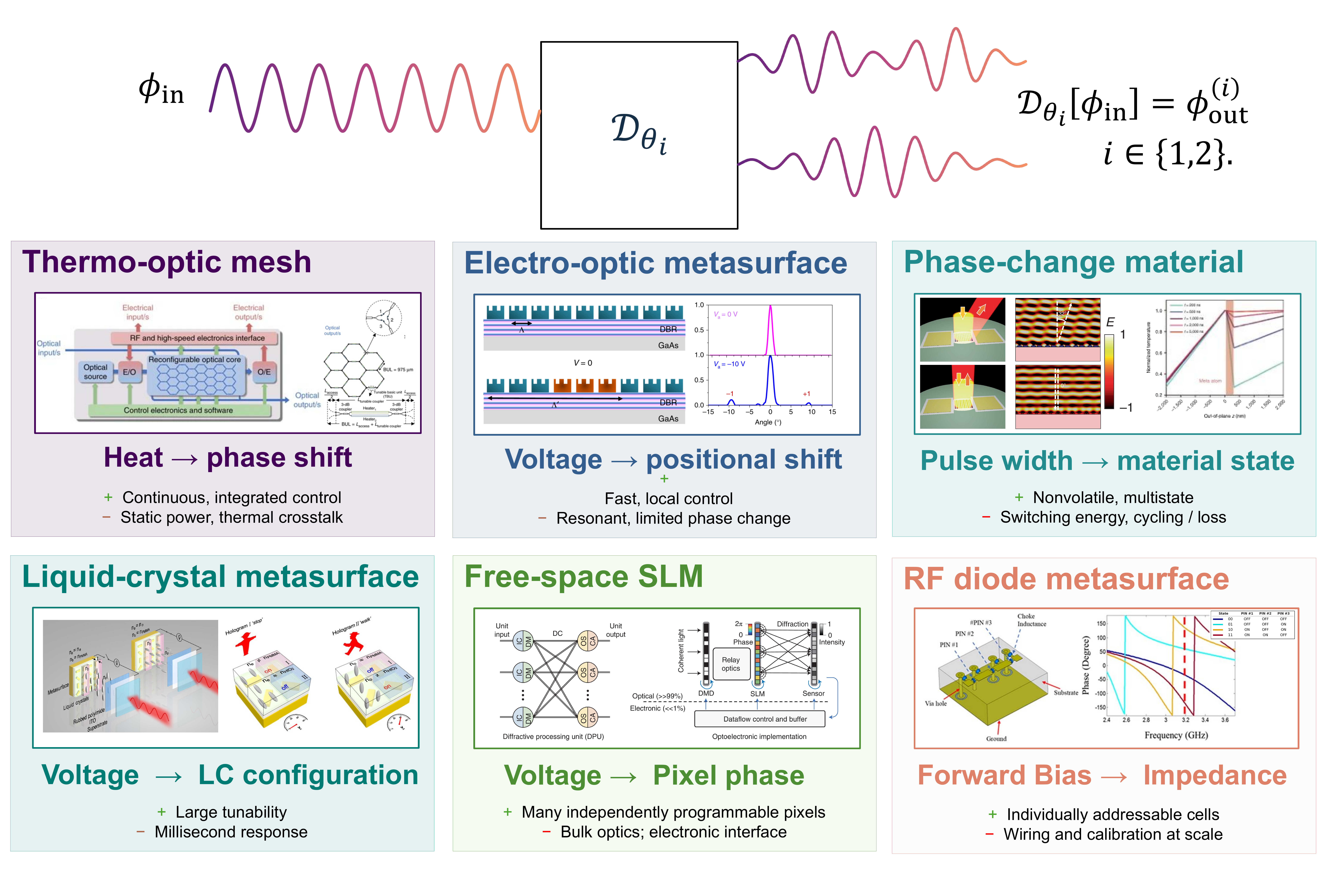}
\caption{
\textbf{Reconfiguration changes the operator implemented by a wave device.}
Adjusting physical parameters changes the output for a fixed input, $D_{\theta_i}[\phi_{\mathrm{in}}]=\phi_{\mathrm{out}}^{(i)}$. Upper row, left to right: thermo-optic meshes \cite{perez2017multipurpose}, electro-optic metasurfaces \cite{wu2019dynamic}, and phase-change metasurfaces \cite{zhang2021electrically}. Lower row: liquid-crystal metasurfaces \cite{li2020electrically}, spatial light modulators \cite{zhou2021large}, and PIN-diode RF metasurfaces \cite{li2019machine}. The control mechanisms impose different trade-offs in tuning range, speed, retention, loss, power, and integration.
\emph{Credits:} Thermo-optic, electro-optic, liquid-crystal,
and RF images reproduced from
refs.~\cite{perez2017multipurpose,wu2019dynamic,li2020electrically,li2019machine},
respectively, under
\href{https://creativecommons.org/licenses/by/4.0/}{CC BY 4.0};
selected portions were resized, arranged, and annotated.
Phase-change and spatial-light-modulator images reproduced
with minor style changes from
refs.~\cite{zhang2021electrically,zhou2021large},
respectively, with permission from Springer Nature Customer
Service Centre GmbH (SNCSC).
The overview schematic and composite layout are original.
}
\label{fig:reconfigurability}
\end{figure*}

The distance between a proof-of-concept structure and a field-programmable computing fabric is set by materials and control. At radio frequencies, loading with varactors or PIN diodes provides fast, low-cost tunability, and the system-level question has shifted to control-plane architecture-how thousands of elements are addressed, calibrated, and networked-as explored in software-defined metasurfaces with embedded controllers~\cite{tsilipakos2020toward,cui2014coding}. Figure~\ref{fig:reconfigurability} summarizes representative reconfiguration technologies. In photonics, the palette is broader and the trade-offs sharper: thermo-optic and electro-optic phase shifters underpin today's programmable circuits~\cite{bogaerts2020programmable}.
In particular, electro-optic control can in principle combine local voltage tuning with a fast response time while reducing heat-mediated crosstalk relative to thermo-optic tuning; electrical and optical coupling may remain \cite{Sinatkas2021}. 
Several routes have been explored, including (i)~transparent conducting oxides (TCOs) such as indium tin oxide (ITO) for electrically controlled carrier-density modulation \cite{Sinatkas2021}. Huang et al. instead model ultrafast all-optical switching in a hybrid ENZ-plasmonic ITO structure; this is not a demonstration of CMOS-voltage tuning \cite{Huang2025enz}. (ii)~Lithium niobate on insulator (LNOI) or organic electro-optic materials have been leveraged for spatial light modulators \cite{Ju2024,Benea2021}. Resonant field enhancement by means of sharp guided-mode resonances or quasi-bound states in the continuum can boost the modest phase shift due to the limited interaction volume \cite{Zografopoulos2023}. (iii)~The voltage-controlled birefringence of liquid crystals combined with dielectric or plasmonic meta-atoms has been exploited in reconfigurable metasurfaces operating with moderate voltages but millisecond (kHz-rate) switching times \cite{Lininger2020}.
Spatial light modulators and digital micromirror devices provide megapixel-scale (if kilohertz-rate) programmability for free-space processors\cite{zhou2021large,pierangeli2019large}.
Chalcogenide phase-change materials add the non-volatility that analog weights demand-zero static power and multilevel storage-demonstrated both in photonic synapses and tensor cores\cite{feldmann2019alloptical,feldmann2021parallel} and in electrically reprogrammable, low-loss phase-change metasurfaces\cite{zhang2021electrically}.

Three cross-cutting lessons emerge from the experimental record. First, calibration and drift, not raw tunability, dominate the engineering budget: architectures that minimize the calibration chain-direct element-wise mappings, self-configuring meshes, adaptive in-system training-consistently translate into better experimental fidelity\cite{tzarouchis2022mathematical,miller2013self,zhou2021large}. Second, non-volatile weight storage is the wave-domain analog of in-memory computing and is likely a precondition for energy claims to survive at system level\cite{feldmann2021parallel,abouhamdan2025programmable}. Third, co-integration with CMOS drivers and converters-rather than the wave physics itself-is where latency and energy are currently won or lost, as the packaged accelerators of Sec.~\ref{sec:integrated} demonstrate\cite{hua2025integrated,ahmed2025universal}.

\subsection{What the experiments demonstrate: scale, precision, speed, and energy}\label{sec:assessment}
An honest synthesis of two decades of demonstrations reads as follows. \emph{Scale:} programmable integrated processors have reached the $10^3$-$10^4$-component regime and 128$\times$128 tensor cores\cite{hua2025integrated,ahmed2025universal}; free-space diffractive systems reach $10^6$ trainable features by spending aperture instead of components\cite{zhou2021large,xu2024taichi}; implicit equation solvers and metastructure computing cores remain at $N$ of order units to tens\cite{estakhri2019inverse,tzarouchis2025programmable,nikkhah2024inversedesigned}. \emph{Precision:} analog wave hardware natively delivers roughly 4-8 effective bits; reaching near-electronic inference accuracy has required architectural error management-adaptive or hardware-in-the-loop training, calibration-light mappings, and hybrid analog-digital partitioning\cite{zhou2021large,ahmed2025universal,momeni2025training}. \emph{Speed:} single-pass latency is picoseconds to nanoseconds and has been demonstrated end to end on chip\cite{ashtiani2022onchip}, with throughput multiplied by wavelength, polarization, and mode parallelism\cite{feldmann2021parallel}. \emph{Energy:} sub-photon-per-multiplication operation demonstrates a low detected optical energy budget for specific inference tasks\cite{wang2022optical}, and large systemic gains have been reported when conversion stages are eliminated\cite{chen2023allanalog}; yet the input/output interface-sources, modulators, detectors, converters-can dominate the budget when these components are included, so honest comparisons with digital accelerators must be made at system level, where the case is promising but not yet closed\cite{mcmahon2023physics,abouhamdan2025programmable,brunner2025roadmap}.

The pattern across these axes is consistent: wave hardware may offer an advantage where the workload is linear-algebra-dense, latency-critical, and tolerant of managed analog precision, provided matched-task system-level measurements include conversion, control, and readout costs-with particular opportunities when the data are already waves, a condition that metadevices are uniquely positioned to exploit.

\subsection{Beyond the state of the art: what metadevices could make possible}\label{sec:beyond}
We close with the scopes that, in our assessment, lie just beyond the demonstrations above-and for which metamaterials, metasurfaces, and metadevices are not merely one option among many, but the natural substrate.

\paragraph{From programmable devices to general-purpose wave processors.}
The DCM machine\cite{tzarouchis2025programmable} and programmable scattering processors\cite{sol2022metaprogrammable,delhougne2018leveraging} prototype a wave-domain analog of the field-programmable gate array: hardware whose mathematical function is written, not fabricated. The missing steps are architectural rather than physical-modular tiling of $N\times N$ kernels into larger operators, cascadable input/output standards, and loss and noise budgets that survive depth-and they parallel, deliberately, the early history of reconfigurable electronics.

\paragraph{Self-designing and self-calibrating hardware.}
Operating a metadevice inside its own design loop\cite{tzarouchis2025programmable}, together with in-physics training methods that no longer require faithful digital twins\cite{momeni2025training,momeni2023backpropagation,xue2024fully,pai2023experimentally}, points to hardware that discovers, maintains, and re-tunes its own function-absorbing fabrication tolerance and environmental drift as training data rather than fighting them as errors. The surrogate- and representation-learning toolchain of Sec.~\ref{sec:math} \cite{pestourie2023physics,marzban2026inverse} then becomes a runtime component of the hardware, not only a design-time one.

\paragraph{Implicit machines: equilibrium, optimization, and control in the physics.}
Feedback metadevices realize fixed points physically\cite{estakhri2019inverse,tzarouchis2022mathematical,cordaro2023solving}, while nonlinear hidden-state dynamics would require a state-dependent material response or nonlinear detection-and-feedback. Nonlinear input encoding instead yields a nonlinear input-output map with linear field propagation at fixed controls\cite{wanjura2024fully,yildirim2024nonlinear}. The natural targets are deep-equilibrium-style inference layers, physical solvers for optimization and model-predictive control with microsecond loop closure, and Ising-type annealers\cite{pierangeli2019large} embedded in metastructures: classes of computation in which relaxation, not arithmetic, is the resource.

\paragraph{Computing-integrated sensing and communications.}
Nowhere is the case stronger than where the data are already electromagnetic waves in flight. The hardware continuum from reconfigurable intelligent surfaces\cite{direnzo2020smart} to programmable computing metasurfaces\cite{liu2022programmable,gao2023programmable,li2019intelligent} suggests 6G-and-beyond infrastructures in which the propagation environment itself filters, classifies, and pre-decodes: over-the-air preprocessing that can avoid digitizing the incident field before the optical or RF operation, while retaining control, detection, and downstream readout costs, integrated sensing and communication with the front-end mathematics performed in the channel, and metadevice equation solvers acting as physical-layer equalizers and estimators. For the antennas and telecommunications community, this reframes the metasurface from a beam-shaping component into a distributed, programmable computer.

\paragraph{Passive optical edge processing and co-designed apertures.} Co-design also allows the optical front end to shape the information passed to downstream computation, optimizing task performance rather than image fidelity alone. For example, meta-optics optimized for a frozen vision-language model improve recognition in simulated imaging benchmarks \cite{kang2026vlmaware}. Privacy-aware meta-optics provide a complementary experimental example, suppressing identity-related image details while retaining information needed for person detection \cite{tasneem2026privacy}.
Passive computing metasurfaces\cite{zhou2020flat,cordaro2023solving} and end-to-end co-designed meta-optics (Sec.~\ref{sec:math})\cite{lin2021endtoend} point to imaging and sensing front ends in which compression, feature extraction, and even classification are implemented by passive optical propagation before detection, without active power in the operator itself but with illumination, detection, and readout energy-the decisive resource allocation for edge devices, space hardware, and always-on sensing.

\paragraph{New physical axes: memory, nonlinearity, and time.}
Non-volatile phase-change meta-atoms\cite{zhang2021electrically,feldmann2019alloptical} give metadevices synaptic memory; structural-nonlinearity schemes give them activation functions without nonlinear materials\cite{yildirim2024nonlinear,wanjura2024fully}; and time-varying media add temporal modulation as a genuinely new computational dimension-parametric gain, nonreciprocal computation graphs, and frequency-synthetic layers within a single structure\cite{galiffi2022photonics}. A metadevice combining memory, activation/nonlinearity, and temporal modulation-with the activation supplied by nonlinear encoding, readout, or hidden-state dynamics as specified by its architecture-would constitute a complete neuromorphic \emph{medium} rather than a neuromorphic component.

\begin{figure*}[t]
\centering
\includegraphics[width=\textwidth]{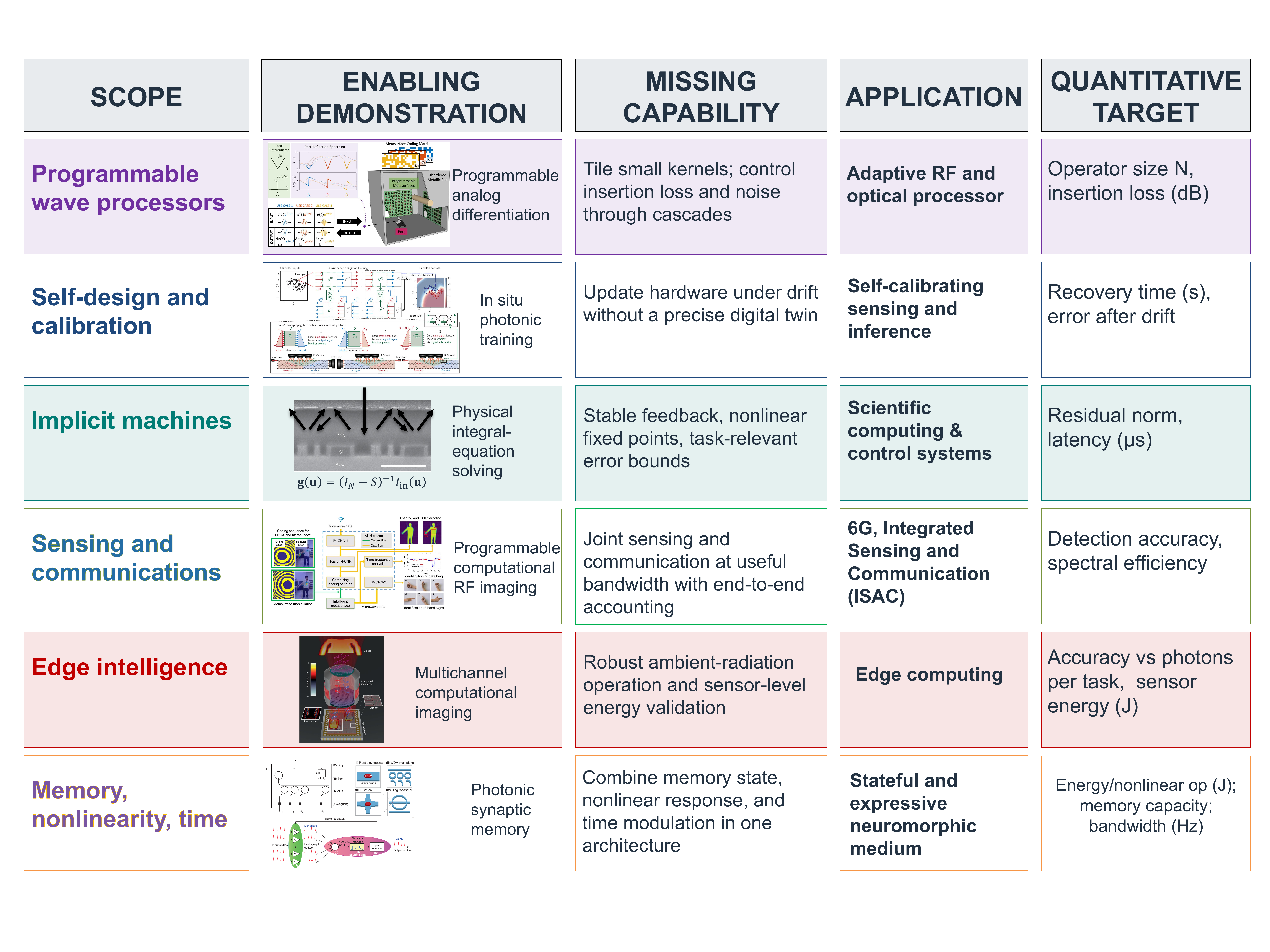}
\caption{
\textbf{From enabling demonstrations to future wave-computing capabilities.}
Each row connects an experimental demonstration to a missing capability, a prospective application, and metrics for evaluating progress. Top to bottom: programmable differentiation \cite{sol2022metaprogrammable}, in-situ photonic training \cite{pai2023experimentally}, free-space integral-equation solving \cite{cordaro2023solving}, intelligent RF imaging \cite{li2019intelligent}, multichannel meta-imaging \cite{zheng2024multichannel}, and photonic synaptic memory \cite{feldmann2019alloptical}. Proposed capabilities and applications are prospective synthesis, not jointly demonstrated systems; metrics identify what to evaluate rather than numerical performance thresholds.
\emph{Credits:} Differentiation and RF-imaging images
reproduced from
refs.~\cite{sol2022metaprogrammable,li2019intelligent},
respectively, under
\href{https://creativecommons.org/licenses/by/4.0/}{CC BY 4.0};
selected portions were resized and arranged.
Training image reproduced/modified with permission from
ref.~\cite{pai2023experimentally}, AAAS.
Equation-solving, meta-imaging, and synaptic-memory images
reproduced with minor style changes from
refs.~\cite{cordaro2023solving} (equation solving), \cite{zheng2024multichannel} (meta-imaging), and \cite{feldmann2019alloptical} (synaptic memory),
respectively, with permission from Springer Nature Customer
Service Centre GmbH (SNCSC).
}

\label{fig:beyond_sota}
\end{figure*}

Figure~\ref{fig:beyond_sota} summarizes these research directions. The arc of the field, in short, runs from materials that compute toward computers that \emph{are} materials: distributed, programmable matter for the wave domain, designed and trained by the mathematical machinery of Secs.~\ref{sec:math} and \ref{sec:sota}, and embedded wherever the world's information already travels as waves.

\section{Outlook and future directions}\label{sec:outlook}

The rapid progress surveyed above establishes that wave-based systems \textit{can} implement mathematical operators, solve equations, and perform increasingly sophisticated neural-network workloads. The more consequential question is where wave-based computing can provide a meaningful system-level advantage. Analog precision remains one important limitation: fabrication disorder, noise, calibration drift, and finite dynamic range constrain the accuracy of physical computation and make performance inherently task dependent \cite{mcmahon2023physics}. At larger scales, input-output conversion may become an even more decisive bottleneck. The advantages of passive propagation and massive optical parallelism can be diminished if substantial energy and latency are required to encode electronic data into optical fields and subsequently detect and digitize the result \cite{abouhamdan2025programmable,xu2024taichi}. Future architectures must therefore  co-design the computational representation, optical hardware, memory, and electronic interfaces rather than optimize the photonic core in isolation.

A second challenge is making these systems adaptive after fabrication. Recent work on physical neural networks suggests that hardware imperfections need not simply be corrected externally, but can instead be incorporated into the learning process through in-situ, forward-only, and backpropagation-free training strategies \cite{momeni2025training,xue2024fully,momeni2023backpropagation}. Such approaches may ultimately allow the physical device itself to participate in optimization and compensate for model mismatch, drift, and component variability. At the same time, the computational scope of metadevices is expanding beyond static linear transformations. Nonlinear computation can emerge from measurement, feedback, recurrence, and appropriately structured wave dynamics, while programmable feedback architectures already demonstrate stationary and non-stationary iterative algorithms including matrix inversion, root finding, and inverse design \cite{wanjura2024fully,tzarouchis2025programmable}. These developments point toward physical implementations of recurrent, dynamical, and implicit computational models rather than optical replicas of conventional feed-forward digital networks.

The earliest practical advantages may consequently arise in applications where information already exists naturally in the wave domain. Optical front ends for imaging and sensing can process spatial information before photodetection, while programmable RF metadevices could perform filtering, inference, or integrated sensing-and-communication directly on incident electromagnetic signals. Edge applications are especially attractive when low latency and reduced data movement matter more than high numerical precision. Determining where such architectures genuinely outperform digital alternatives, however, will require standardized benchmarks that report not only throughput and energy efficiency but also precision, latency, calibration cost, and the boundaries of what is included in system-level measurements \cite{brunner2025roadmap,mcmahon2023physics}. Ultimately, the most distinctive opportunity for neuromorphic nanophotonics may not be in reproducing digital computation with light, but in co-designing algorithms with propagation, scattering, feedback, and learning so that the physics itself becomes the computational process.

\section*{Acknowledgements}
X. Hadia and R. Pestourie acknowledge support from the U.S. National Science Foundation under Grant No. IIS 2435905 and the National Institute of Biomedical Imaging and Bioengineering of the National Institutes of Health under Award Number R21EB036343.
O. Tsilipakos acknowledges partial support from the TPCI/NHRF internal project PHOTOPMETA. 

\section*{Author contributions}
Conceptualization: D.C.T.; writing of Secs.~\ref{sec:math} and
\ref{sec:sota}: X.H., D.C.T., and R.P.; writing of Sec.~\ref{sec:devices}:
D.C.T. and O.T.; writing of Secs.~\ref{sec:intro} and~\ref{sec:outlook}, review and editing: all authors; supervision: D.C.T.\ and R.P.;
submission and correspondence: R.P.\ and D.C.T.\

\section*{Data availability}
Data sharing is not applicable to this article as no datasets were generated or analyzed during the current study. All literature data cited in Figure~\ref{fig:performance_landscape} and Table~\ref{tab:fig4_benchmarks} are derived from published articles referenced in the text.

\section*{Code availability}
The plotting and analysis scripts used to generate the benchmark comparison in Figure~\ref{fig:performance_landscape} are available from the corresponding authors upon reasonable request.

\section*{Competing interests}
D.C.T.\ is a Senior Research Engineer with Intracom Telecom S.A.

\bibliographystyle{unsrtnat} 
\bibliography{references}

\end{document}

%% file: figures/fig4_benchmarks.tex
\begin{table}[p]
\centering
\caption{\textbf{Coordinate provenance and comparison limits for Figure~\ref{fig:performance_landscape}.} Each pair is throughput (TOPS) followed by efficiency (TOPS/W). Digital TFLOPS are expressed numerically in TOPS; optical entries retain their authors' equivalent-operation definitions.}
\label{tab:fig4_benchmarks}
\footnotesize
\setlength{\tabcolsep}{5pt}
\renewcommand{\arraystretch}{1.35}
\begin{tabularx}{\textwidth}{@{}>{\raggedright\arraybackslash}p{2.2cm}>{\raggedright\arraybackslash}p{2.5cm}>{\raggedright\arraybackslash}X>{\raggedright\arraybackslash}X@{}}
\toprule
Platform & TOPS; TOPS/W & Operations and workload & Power boundary and evidence \\
\midrule
A100 SXM\cite{nvidiaA100specs2026} & 312; 0.78 & Dense FP16/BF16 Tensor Core peak; two floating-point operations per MAC. & 400~W board TDP; vendor peak/TDP, not measured application efficiency. \\ \specialrule{0.2pt}{5pt}{5pt}
H100 / H200 SXM\cite{nvidiaH100specs2026,nvidia2024h200} & 989.5; 1.41357 & Same dense FP16/BF16 convention; both GPUs have identical plotted coordinates. & 700~W maximum board TDP/GPU; excludes host and facility power. \\ \specialrule{0.2pt}{5pt}{5pt}
PACE\cite{hua2025integrated} & 8.19; 2.38 & Paper-defined MAC throughput; recurrent accelerator demonstrations. Mean 7.61 ENOB in characterization is not FP8 equivalence. & Reported timing/energy including lasers; excludes the 4.21~TOPS/W laser-free value. Full wall-plug boundary unverified. \\ \specialrule{0.2pt}{5pt}{5pt}
S$^2$NN\cite{xu2026sparse} & 417.96; 12.589 & Four-layer MNIST configuration; complex optical transformations expanded into equivalent real operations. Binary output and 8-bit SLM phase settings do not establish arithmetic precision. & 33.2~W component budget: DMD 11.2, sensor 0.5, SLM 11.2, laser 5, threshold light 5.3~W; efficiency from equivalent throughput/power. \\ \specialrule{0.2pt}{5pt}{5pt}
Hu 2025, integrated TFLN\cite{hu2025integrated} & 0.01033; 0.11 & Demonstrated single-channel random-vector multiplication at 10.33~GOPS; authors' operation convention. This is not the inferred 31-channel headline configuration. & Estimated 9.015~pJ/OP (Table~S2): DFB drive, heater, modulation, detector; assumed DAC, memory and ADC costs. Source-rounded efficiency. \\ \specialrule{0.2pt}{5pt}{5pt}
Wang 2024, NG-RC\cite{wang2024ultrafast} & 211; 41 & Single wavelength, 60~Gbaud; eight delays and 45 outputs. Complex reservoir operations expanded into real equivalents. & Estimated 5.1~W (Note~5): laser, EDFA, DAC, modulation, heaters, ADC and detector. Source-rounded efficiency. \\ \specialrule{0.2pt}{5pt}{5pt}
ACCEL\cite{chen2023allanalog} & 4,550; 74,800 & Methods values for three-class ImageNet. Task-specific optical/analog equivalent count, with more than 99\% assigned to optics; not a general GEMM benchmark. & Reported task timing/energy: illumination, SRAM, control and analog electronics included; full host/facility boundary unverified. \\
\bottomrule
\end{tabularx}
\par\vspace{5pt}
\begin{minipage}{\textwidth}
\footnotesize
\textit{Digital derivation.} Dense throughput is half the rounded advertised sparse FP16/BF16 peak: $624/2=312$ for A100 and $1979/2=989.5\simeq990$ for H100/H200. Efficiency is dense peak divided by board TDP. The derived decimals do not imply additional source precision. The comparison controls operand format, sparsity, device count and power-boundary type; it does not measure matched application energy.\par\smallskip
\textit{Optical interpretation.} Optical entries are retained as source-defined literature metrics with illumination terms included. Their precision, tasks, equivalent-operation conventions and peripheral boundaries are not normalized to one another or to the GPU entries; ratios across these reporting bases do not demonstrate application-level advantage.\par\smallskip
\textit{Excluded entries.} DGX B200 uses an eight-GPU whole-server boundary; estimates without an identified primary source or combining different configurations are omitted; TeMPO has an unresolved throughput/power/efficiency inconsistency in a modeled architecture; FAST's throughput and energy model use different channel configurations and apportioned SLM power.
\end{minipage}
\end{table}